\documentclass[]{article}  
\usepackage[margin=25mm]{geometry}
\usepackage{authblk} 
\usepackage{amsmath,amsfonts,amssymb}
\usepackage{siunitx}
\usepackage{graphicx}
\providecommand{\keywords}[1]
{
  \small	
  \textbf{\textit{Keywords---}} #1
}

\usepackage{hyperref}
\hypersetup{colorlinks,allcolors=blue}
\usepackage{cleveref}
\title{Modeling light-controlled actuation of flexible magnetic composite structures using the finite element method (FEM)}
\author[a,c]{Amit Kumar Jha}
\author[b]{Meng Li}
\author[c]{Ewan S. Douglas}
\author[c]{Erin R. Maier}
\author[b]{Fiorenzo G. Omenetto}
\author[d]{Corey Fucetola}
\affil[a]{Wyant College of Optical Sciences, University of Arizona, 1630 E. University Blvd., Tucson, AZ - 85719, USA}
\affil[b]{Silklab, Department of Biomedical Engineering, Tufts University, 4 Colby Street, Medford, MA - 02155, USA}
\affil[c]{University of Arizona Space Astrophysics Lab, Steward Observatory, University of Arizona, 933 N. Cherry Avenue, Tucson, AZ - 85721, USA}
\affil[d]{Sublamit Laboratories, 46 Conwell Avenue, Somerville, MA - 02144, USA}
\date{}

\begin{document} 
\maketitle
\begin{abstract}
Photoactive materials hold great promise for a variety of applications. We present a finite element model of light-controlled flexible magnetic composite structure composed of 33.3\% Chromium dioxide ($\text{CrO}_{\text{2}}$) and 66.7\% Polydimethylsiloxane (PDMS) by weight. The structure has a dimension of 8 \si{mm} x 2 \si{mm} x 100 \si{\mu m} and has been previously experimentally studied. Due to the low Curie temperature, the structure acts as an actuator, shows significant deflection under the external magnetic field and relaxation due to laser heating. Thermal and magnetic deflection analysis has been performed using the FEM model. The simulation results show a maximum structural deflection of 6.08 \si{mm} (76\% of the length of the structure) when subjected to 30 \si{mT} magnetic flux density and 160 \si{mW} laser power at 303 \si{K} (room temperature). We will present the results of the simulation model and comparison to experimental data reproducing the previously observed motion of the ($\text{CrO}_{\text{2}}$+PDMS). This model will enable future fracture and fatigue analysis as well as extension to new photoactive geometries.
\end{abstract}


\section*{Introduction}
\label{sec:intro}  
The interest of researchers in actuators and actuator based components have grown rapidly over the past decade. These are devices that move or deform in response to stimuli. Electromagnetic sources and mechanical forces are some common stimuli that have been used to drive these devices and their application can be seen in the fields of soft-robotics, stretchable electronics, and optomechanics~\cite{kramer2015soft}. Among these stimuli, also comes light which offers the advantage of contactless control~\cite{kwan2018light} and localized stimulation in actuation structures. Examples include photomechanical systems~\cite{ritter2004novel} like optical tweezers~\cite{van2010optomechanical}, optical fibers~\cite{ye2012azobenzene}, and gradient material structures~\cite{zhang2014photoactuators} where light-sensitive materials are used to achieve the actuation function.
\\
Magnetic actuators~\cite{xu2015self,keyes2009computational} are also widely studied because of their applications in the area of optomechanics and micromechanics. They are generally made of either microscopic magnetic beads~\cite{keyes2009computational} arrayed in between sheets of polymers or doping a polymer with ferromagnetic~\cite{xu2015self} materials. A major disadvantage of these ferromagnetic actuators is that they have high Curie temperature which restricts these actuators to be used to their full potential. At the Curie temperature, a ferromagnetic material becomes paramagnetic which provides a great scope for using them as actuator materials. However, due to their high Curie temperature (roughly around 600-1200 \si{K}), full actuation can't be achieved at a convenient temperature range.
\\
Recently, Meng Li et al.~\cite{li2018flexible} have demonstrated full actuation control using a composite polymer made of 33.3\% $\text{CrO}_\text{2}$ and 66.7\% PDMS by weight. This was achieved due to the low Curie temperature (around 395 \si{K}) of the treated $\text{CrO}_\text{2}$, allowing the composite structure to realize full actuation function. The composite polymer has a gradient structure and an overall dimension of 8 \si{mm} x 2 \si{mm} x 100 \si{\mu m}. A permanent magnetic field source was used to deliver the load on the composite structure causing the deflection. On the other hand, a laser source was used to heat the surface causing demagnetization of the ($\text{CrO}_\text{2}$+PDMS) composite resulting in almost full relaxation.
\\
In this work, we present a finite element model of the same composite structure and compare the simulation results of our model with that of the experimental results. We present the deflection analysis of the composite structure due to the magnetic source and also the demagnetization effects on the deflection due to laser heating. The FEA model provides a platform to study these structures and one can simulate both simple and intricate geometries of these composite polymers using our simulation model. In Section~\ref{sec:1}, we will describe our modeling approach to develop the FEM model whereas in section~\ref{sec:3}, we will show the simulation results and compare it with the experimentally obtained data. Section~\ref{sec:4} will summarize the main findings of this work.
\section{Modeling}
\label{sec:1}
In this section, we describe our approach to developing the FEM model and defining the necessary multiphysics definitions. We have used the COMSOL Multiphysics\textsuperscript{\textregistered} software suite~\cite{COMSOL} to perform the Finite Element Analysis (FEA) study of our composite structure model. COMSOL was chosen because we are using the structural mechanics, heat transfer in solids,  and AC/DC multiphysics modules to simulate and add the multiphysics definitions to our desired geometry.
In addition to that, MATLAB\textsuperscript{\textregistered} software~\cite{MATLAB} has also been used for post processing of our simulation data. 
\subsection{Geometry}
The geometry of our model is a cantilever beam structure with an overall dimension of 8 \si{mm} x 2 \si{mm} x 100 \si{\mu m}. The structure is composed of two layers namely P-Side and C-Side. The former layer represents cross-linkage PDMS polymer with almost no concentration of $\text{CrO}_\text{2}$. The latter, on the other hand, represents the cross-linkage PDMS with $\text{CrO}_\text{2}$ concentration. Both the layers have an equal thickness of 50 \si{\mu m }. The P-Side is transparent whereas the C-Side is opaque due to the presence of $\text{CrO}_\text{2}$. 
The composite polymer structure used by Meng Li et al.~\cite{li2018flexible} has the same overall dimension and distinct gradient structure also referred to as P-Side and C-Side layers. Hence, for the simulation purpose, we have considered the same approach and assigned equal thickness to the layers.
   \begin{figure} [ht]
   \begin{center}
   \begin{tabular}{c} 
   \includegraphics[scale = 0.47]{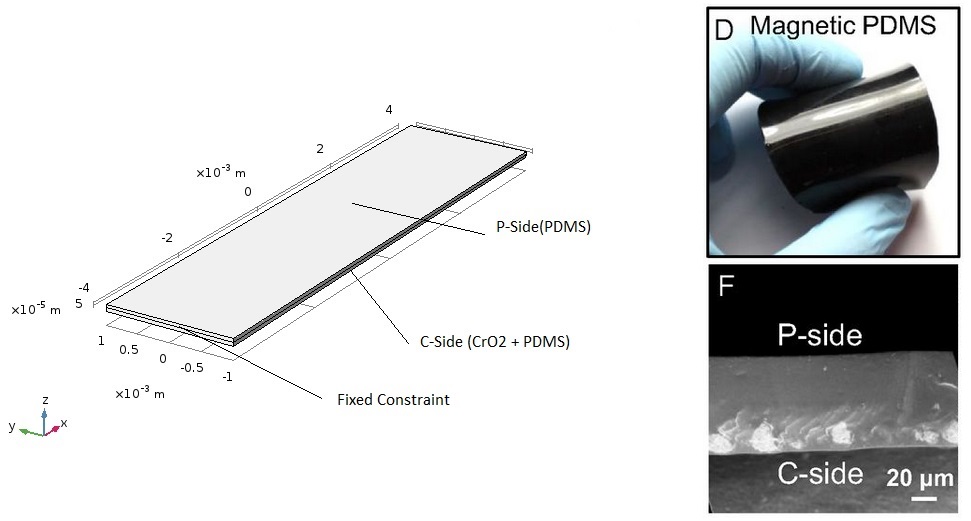}
   \end{tabular}
   \end{center}
   \caption[1] 
   { \label{fig:1} 
Cantilever beam geometry of the composite structure (left). Both the layers P-Side (top layer) and C-Side (bottom layer) have an equal dimension of 8 \si{mm} x 2 \si{mm} x 50 \si{\mu m}. Images of the composite structure (right) taken from fig 1D and 1F of Meng Li et al.~\cite{li2018flexible}}
   \end{figure} 
\subsection{Material properties}
The following are the experimentally determined material properties that have been used to simulate our model and the composite layers:  
\begin{table}[ht]
\caption{Material properties of P-Side (crosslinkage PDMS polymer) and C-Side (crosslinkage PDMS with 33.3\% $\text{CrO}_\text{2}$) layers of the composite structure.}
\begin{center}
\label{tab:mat}
\begin{tabular}{|l|l|l|}
\hline
\rule[-1ex]{0pt}{3.5ex}  \textbf{Material properties} & \textbf{P-Side(PDMS)} & \textbf{C-Side($\text{CrO}_\text{2}$+PDMS)}\\
\hline
\hline
\rule[-1ex]{0pt}{3.5ex}  Young's Modulus & 2.13[\si{MPa}] & 2.23[\si{MPa}]\\
\hline
\rule[-1ex]{0pt}{3.5ex}  Poisson's Ratio & 0.49 & 0.45\\
\hline
\rule[-1ex]{0pt}{3.5ex}  Density & 0.96[\si{g/cm^{3}}] & 1.4[\si{g/cm^{3}}]\\
\hline
\rule[-1ex]{0pt}{3.5ex}  Thermal Conductivity & 0.2[\si{W/(m \cdot K)}] & 0.25[\si{W/(m \cdot K)}]\\
\hline
\rule[-1ex]{0pt}{3.5ex}  Heat Capacity at Constant Pressure & 2174[\si{J/(kg \cdot K)}] & 1840[\si{J/(kg \cdot K)}] \\
\hline
\rule[-1ex]{0pt}{3.5ex}  Co-efficient of Thermal Expansion & $1.88 \times 10^{-4}$[\si{1/K}] & $1.588 \times 10^{-4}$[\si{1/K}] \\
\hline
\end{tabular}
\end{center}
\end{table} 
\subsection{Magnetic properties of the $(\text{CrO}_\text{2}+\text{PDMS})$ composite}
\begin{figure} [ht]
   \begin{center}
   \begin{tabular}{c} 
   \includegraphics[scale = 0.47]{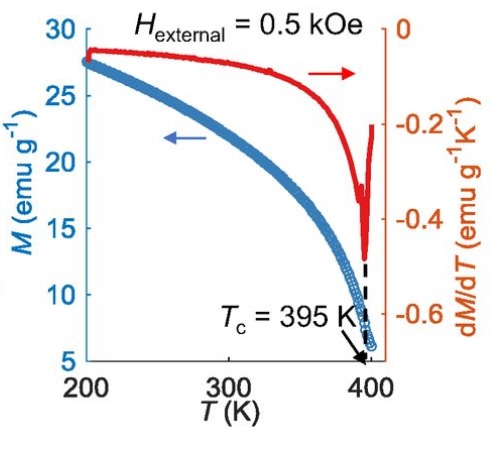}
   \includegraphics[scale = 0.47]{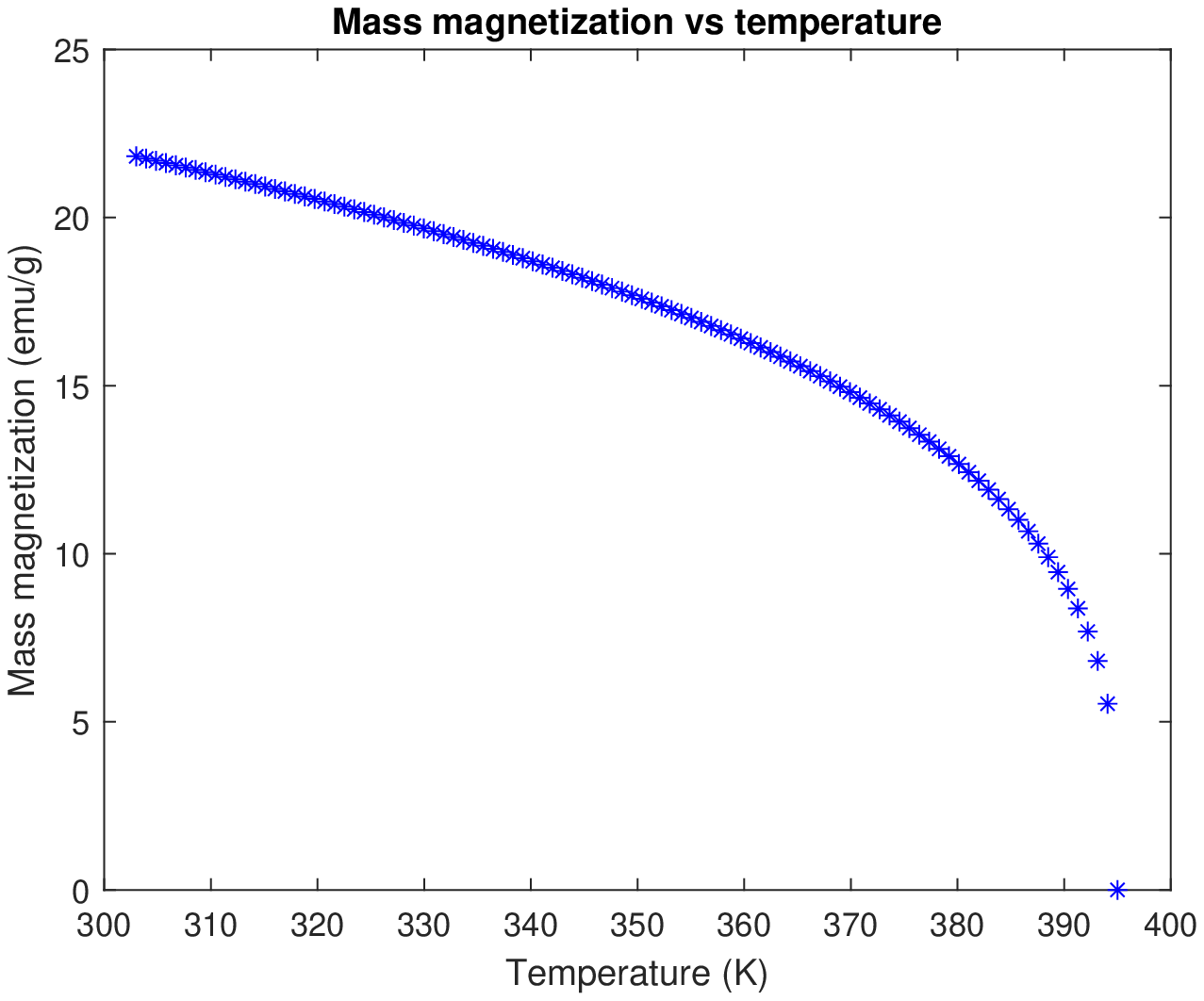}
   \end{tabular}
   \end{center}
   \caption[2] 
   { \label{fig:2} 
   Plot of mass magnetization vs temperature taken from fig 2D of Meng Li et al.~\cite{li2018flexible}(left). Plot of mass magnetization vs temperature governed by eq~\ref{eq:1} (right). The relationship describes the decreasing magnetization properties of the composite structure with increasing temperature. The magnetization comes to zero when the temperature is equal to the Curie temperature.}
   \end{figure}

The magnetic properties of the $\text{CrO}_\text{2}$+PDMS composite has been experimentally measured and is detailed in the works of Meng Li et al~\cite{li2018flexible}. The plot shown in fig~\ref{fig:2} (left) shows the relationship between the mass magnetization and the temperature of the composite structure. On fitting a curve, we can obtain the following relation between mass magnetization $|\vec m(T)|$ and the temperature $T$ :
\begin{equation}
\label{eq:1}
    |\vec m(T)| = C(T_{C}-T)^\beta
\end{equation}
where $C = 5.661$ is the Curie Constant and $\beta = 0.2984$ is the critical exponent factor respectively. The Curie temperature of the treated $\text{CrO}_\text{2}$ is 395 \si{K} and is expressed as $\text{T}_\text{C}$ in eq~\ref{eq:1}.
Also, we can observe that the magnetization properties are temperature-dependent and as the temperature increases, the magnetization decreases. Hence, we can control the magnetization properties by altering the temperature of the composite structure.
\subsection{Loading due to a permanent magnetic field source}
In the works of Meng Li et al.~\cite{li2018flexible}, it has been mentioned that the composite structure is subjected to a permanent magnetic field source to create the deflection. The value of the magnetic flux density at the P-Side of the composite structure is 30 \si{mT}. Using the COMSOL AC/DC module, we have simulated a magnetic field using the properties of a Neodymium magnet manufactured by KJ Magnetics~\cite{KJmag} to realize the same magnetic flux density at our model geometry. The following table has the specifications and the magnetic properties of the permanent magnet used in our simulation model.
\begin{table}[ht]
\caption{Specifications and magnetic properties of the Neodymium magnet used in our model.}
\begin{center}
\label{tab:kjspec}
\begin{tabular}{|l|l|}
\hline
\rule[-1ex]{0pt}{3.5ex}  \textbf{Specification} & \textbf{Neodymium Magnet}\\
\hline
\hline
\rule[-1ex]{0pt}{3.5ex}  Model & DC2E\\
\hline
\rule[-1ex]{0pt}{3.5ex}  Geometry Type & Cylindrical(Disc)\\
\hline
\rule[-1ex]{0pt}{3.5ex}  Radius & 0.375[\si{inch}]\\
\hline
\rule[-1ex]{0pt}{3.5ex}  Height & 0.125[\si{inch}]\\
\hline
\rule[-1ex]{0pt}{3.5ex}  Remnant Flux Density & 13200[\si{Gauss}]\\
\hline
\rule[-1ex]{0pt}{3.5ex} Relative Permeability & 1.01\\
\hline
\end{tabular}
\end{center}
\end{table} 
Due to the magnetic field, the composite structure experiences a deflection load which is governed by the following equation :
\begin{equation}
\label{eq:2}
\vec F(\vec r,t) = (\vec m\cdot\vec\nabla)\vec B(\vec r,t)
\end{equation}
where $\vec F(\vec r,t)$ is the force acting on the composite structure, $\vec m$ is the magnetic moment, and $\vec B(\vec r,t)$ is the magnetic field density. The above equation
is the special case of electromagnetic force density equation~\cite{mansuripur2013force,mansuripur2017field}. Since, for our simulation purpose, we have magnetic field as the only source, the effects of other fields and sources are ignored and the electromagnetic force density equation simplifies to eq~\ref{eq:2}. Equation~\ref{eq:2} is the governing equation for our structural deflection analysis due to the effects of magnetic field.
\subsection{Laser heating of the composite structure}
To see the demagnetization effects, a laser source is required to heat the surface from the C-Side of the composite polymer. C-Side, being opaque, will absorb the incident heat from the laser and will cause a change in the surface temperature. An increase in temperature will result in demagnetization which is governed by eq~\ref{eq:1} causing relaxation which is governed by eq~\ref{eq:2}. Using the deposited beam power feature in COMSOL's heat transfer in solids module, we can implement a laser heating source at the C-Side of our model geometry. The source that we are using has a gaussian distribution and is governed by the equation :
\begin{equation}
\label{eq:4}
{P}_{Source}(r) =  {P}_{Laser}(\frac{1}{2\pi \sigma^2}exp(-\frac{r^2}{2\sigma^2}))
\end{equation}
where $\text{P}_\text{Laser}$ is the laser power and $\sigma = 1$ $\si{cm}$. Also, we will require the heat transfer co-efficient, absorption co-efficient, and the ambient temperature values for implementing the convective heat loss in our model. The heat transfer co-efficient for our model is 49.21 $\si{W/m^2.K}$ whereas the absorption co-efficient is 0.97, and ambient temperature is 303 $\si{K}$ (room temperature) respectively.
   \begin{figure} [ht]
   \begin{center}
   \begin{tabular}{c} 
   \includegraphics[scale = 0.41]{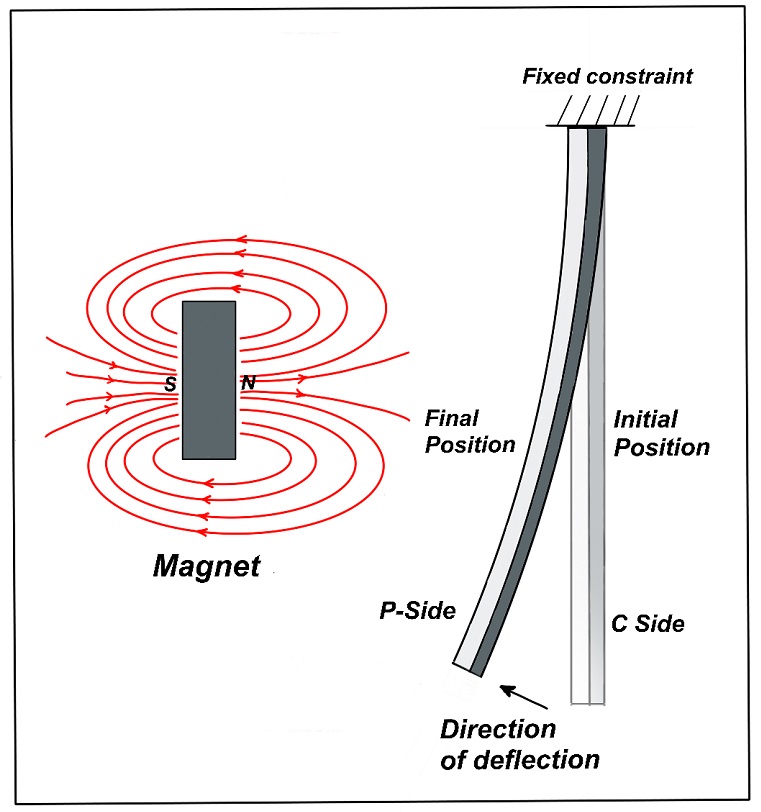}
   \includegraphics[scale = 0.4]{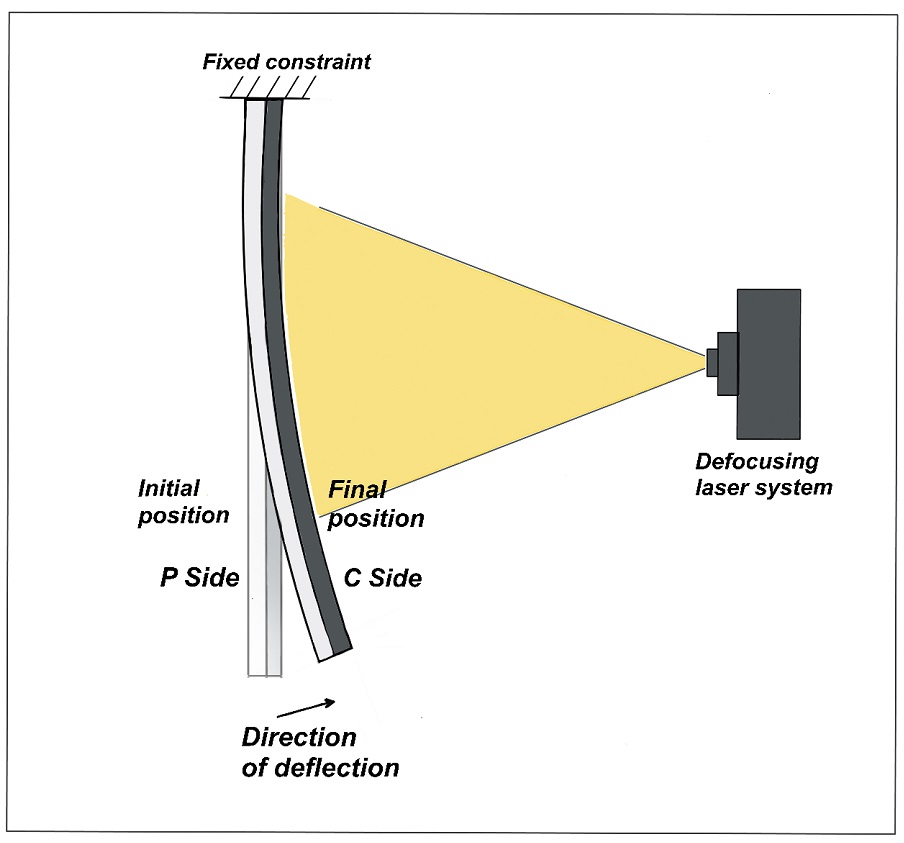}
   \\
   \includegraphics[scale = 0.4]{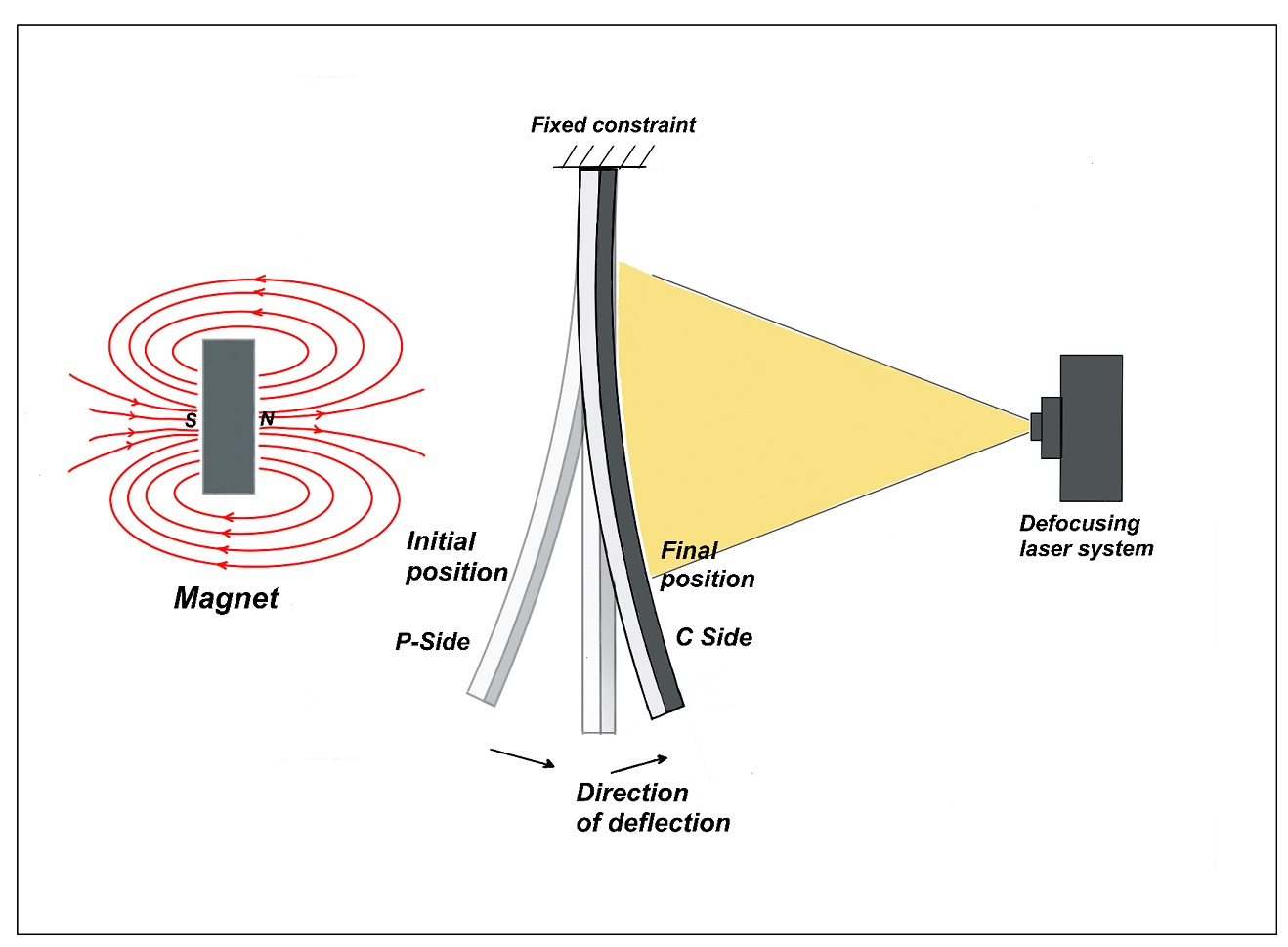}
   \end{tabular}
   \end{center}
   \caption[d1] 
   { \label{fig:d1} 
Illustration~\cite{AUTODESK} showing the direction of deflection of the composite structure when subjected to different stimuli. The direction of deflection is towards the P-Side when only magnetic load is applied on the P-Side (top left). Deflection is towards the C-Side when only laser heating is subjected on the C-Side (top right). Direction of total deflection due to demagnetization and laser heating is towards the C-Side (bottom center).}
   \end{figure} 
\section{Simulation Results}
\label{sec:3}
   \begin{figure}[p]
   \begin{center}
   \begin{tabular}{c} 
   \includegraphics[scale = 0.55]{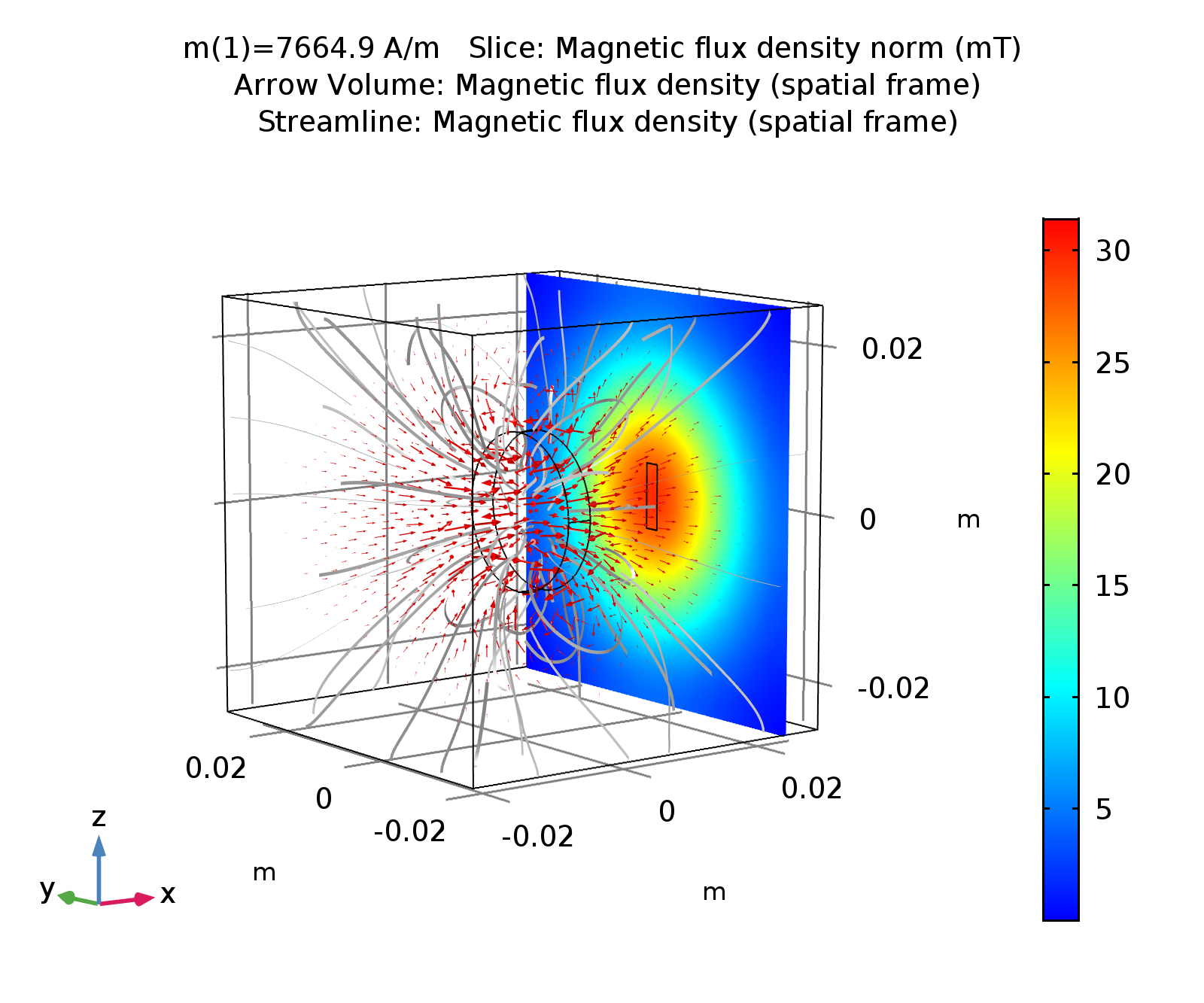}
     \includegraphics[scale = 0.55]{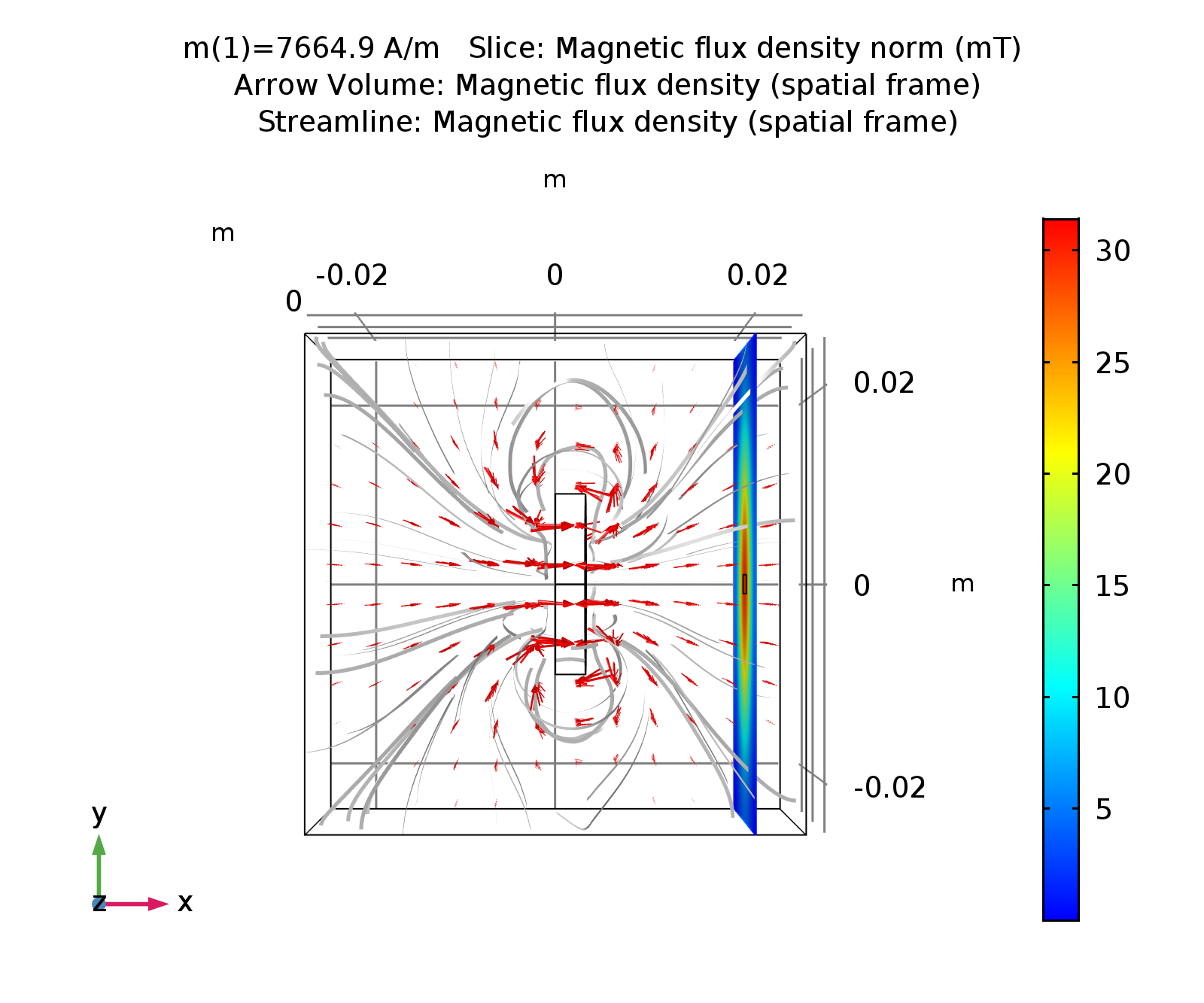}
     \\
       \includegraphics[scale = 0.55]{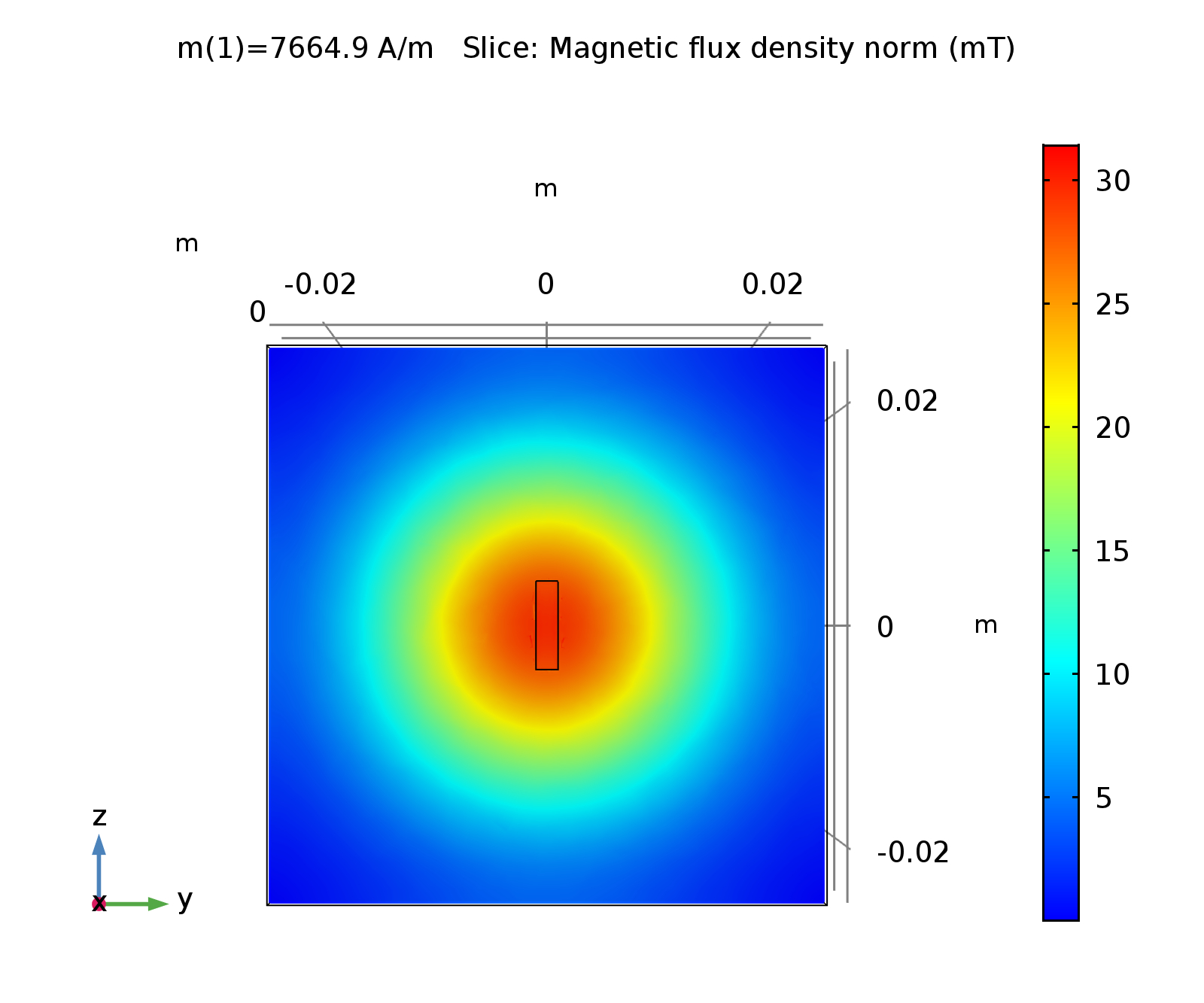}
       \includegraphics[scale = 0.55]{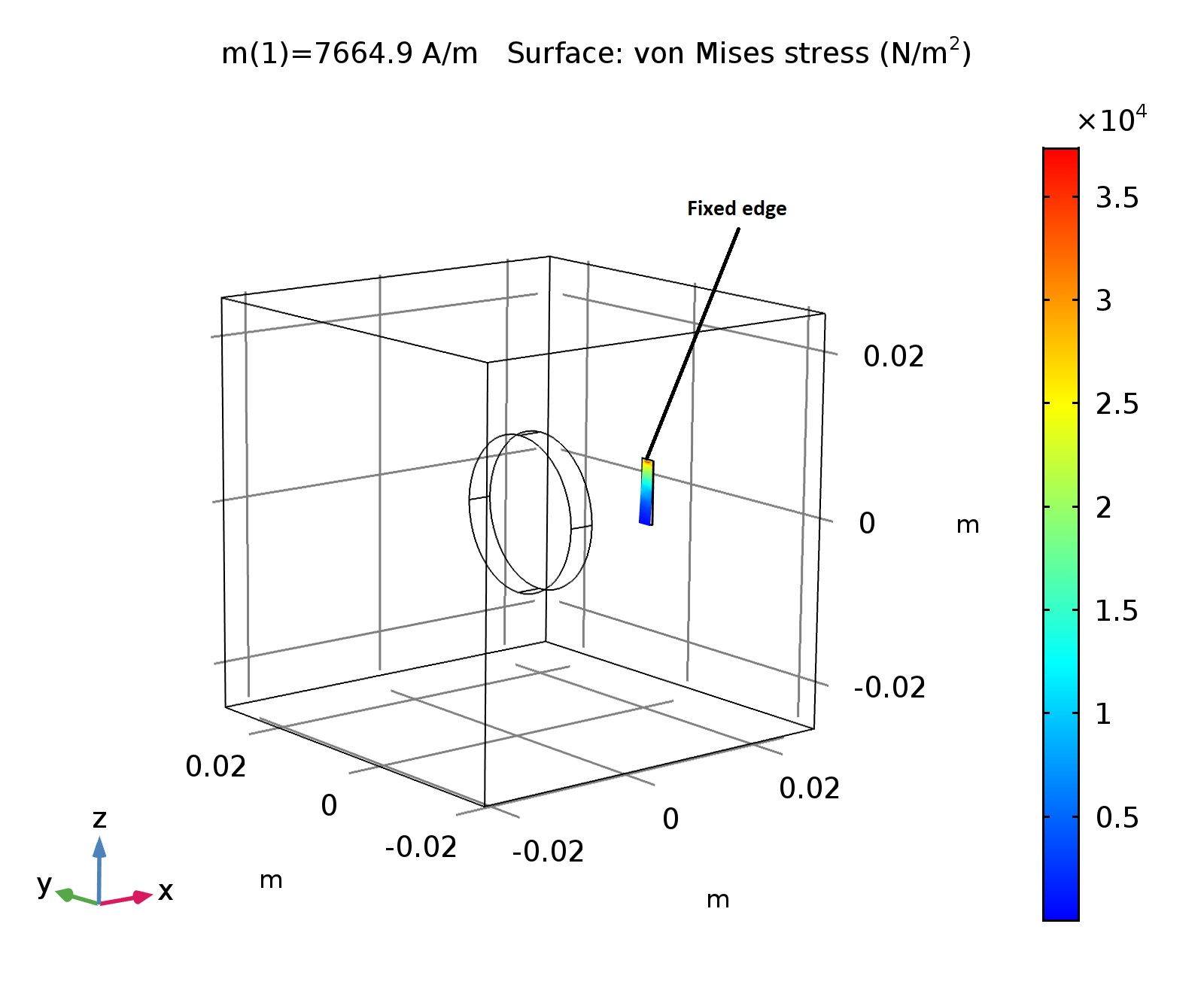}
       \\
        \includegraphics[scale = 0.55]{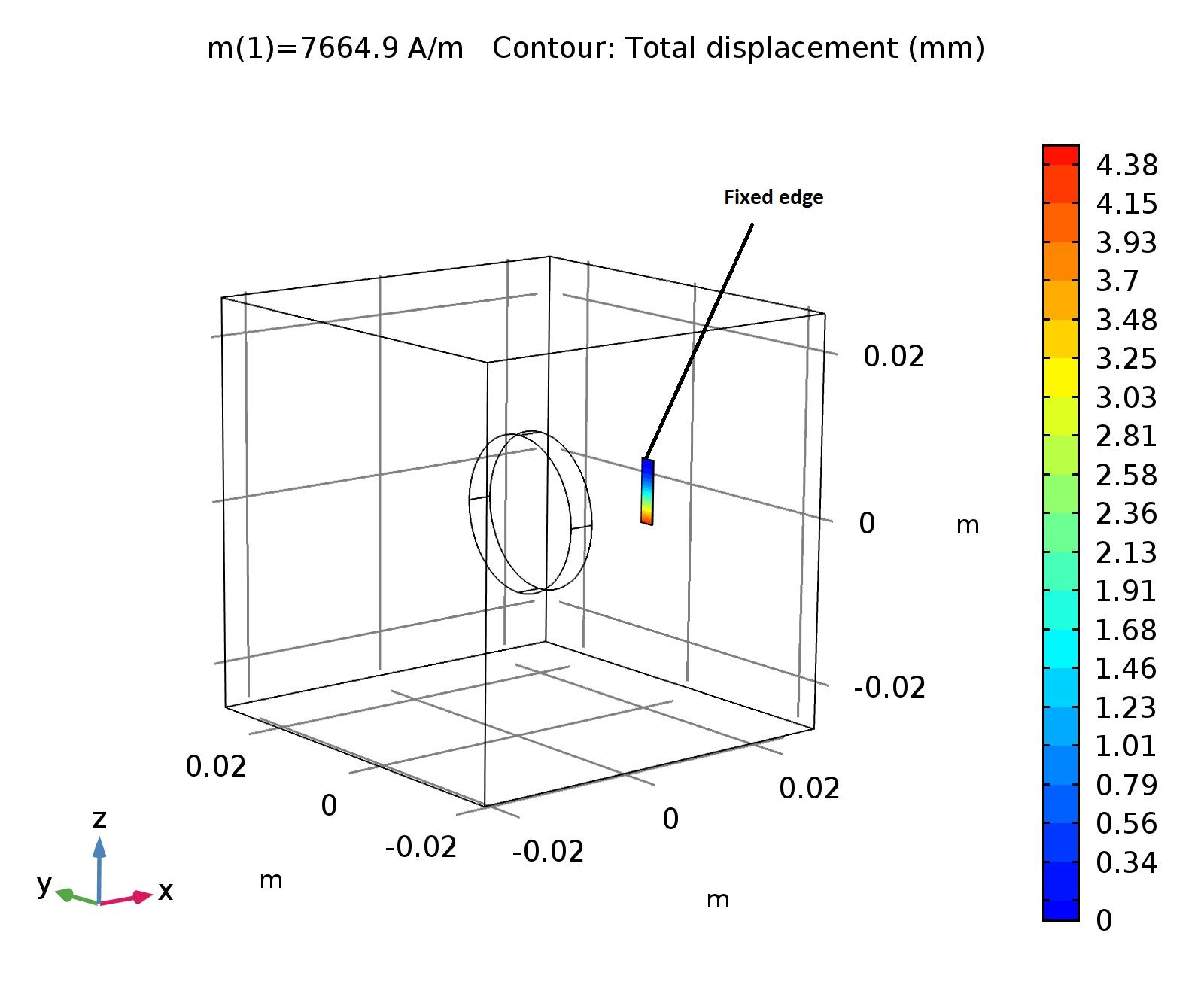}
   \end{tabular}
   \end{center}
   \caption[3] 
   { \label{fig:3} 
Magnetic field lines and magnetic flux density in 3D (top left) and 2D (top right). Magnetic flux density of 30 \si{mT} at the P-Side of the composite geometry (middle left). von Mises stress profile across the composite geometry when the laser is off (middle right). Deflection due to magnetic load (bottom). The deflection is towards the magnet and the P-Side is facing the magnet. The separation between the magnet and the composite structure is 0.02 \si{m}.}
\end{figure} 
   \begin{figure} [ht]
   \begin{center}
   \begin{tabular}{c} 
   \includegraphics[scale = 0.57]{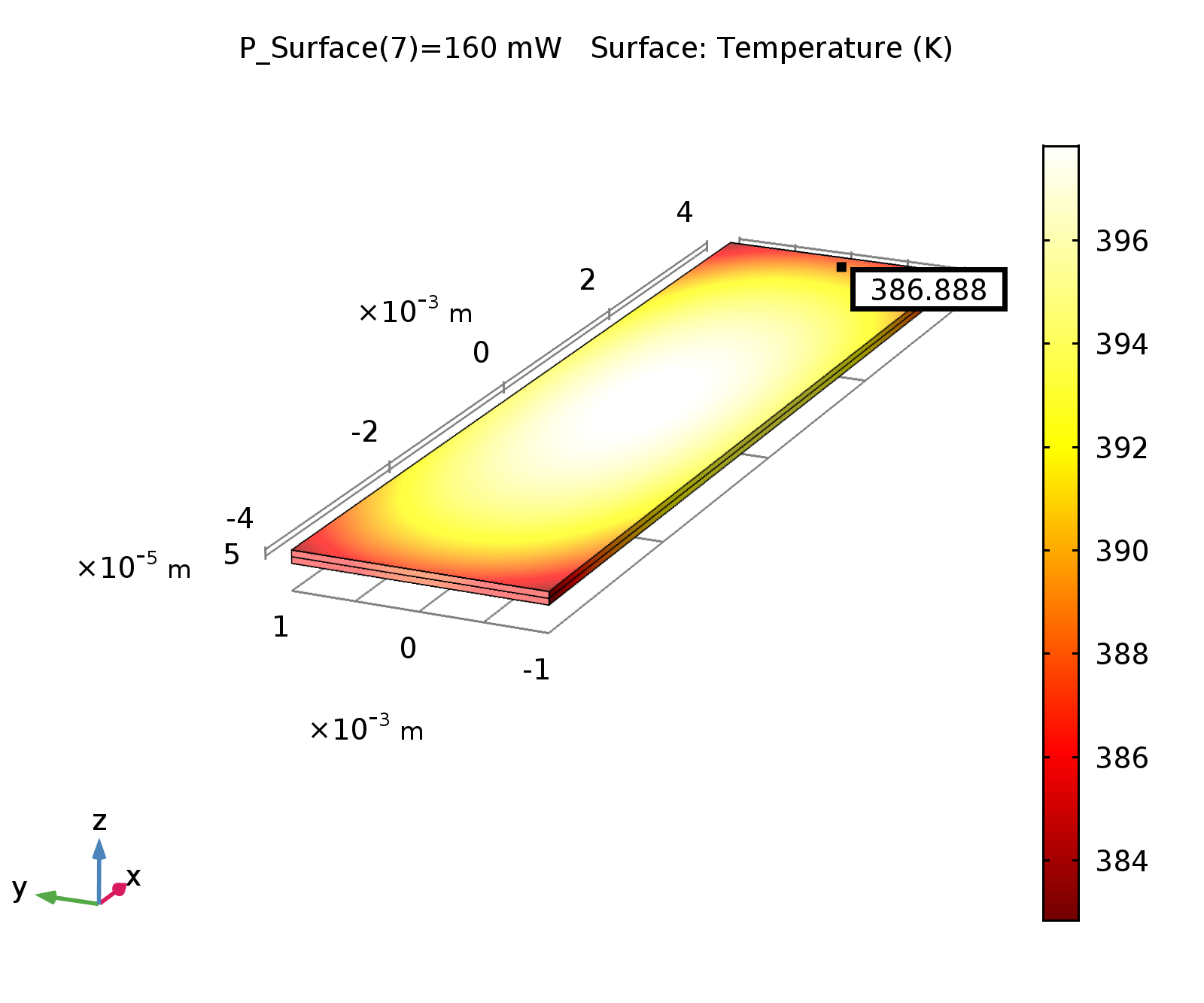}
     \includegraphics[scale = 0.57]{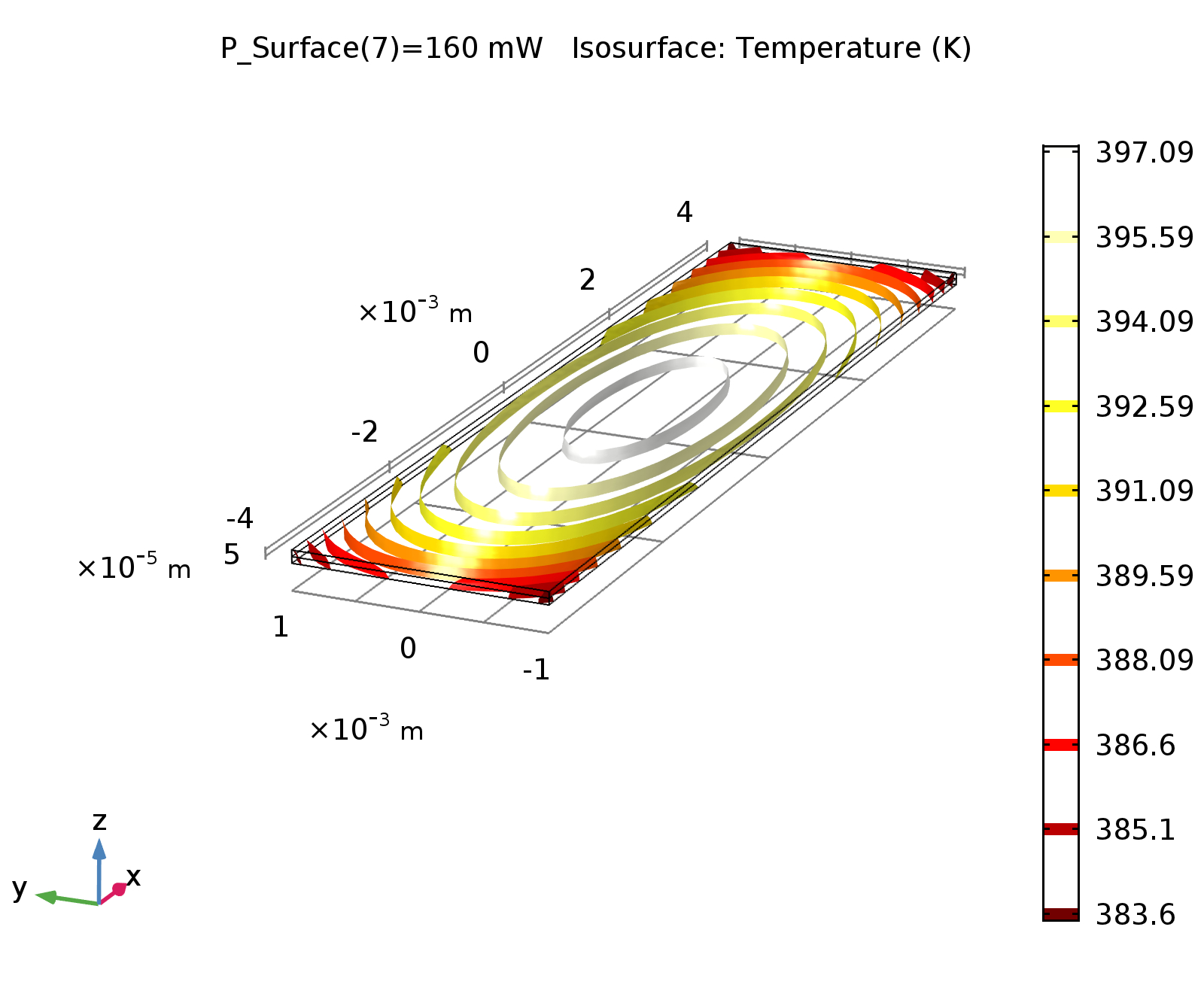}
     \\
       \includegraphics[scale = 0.57]{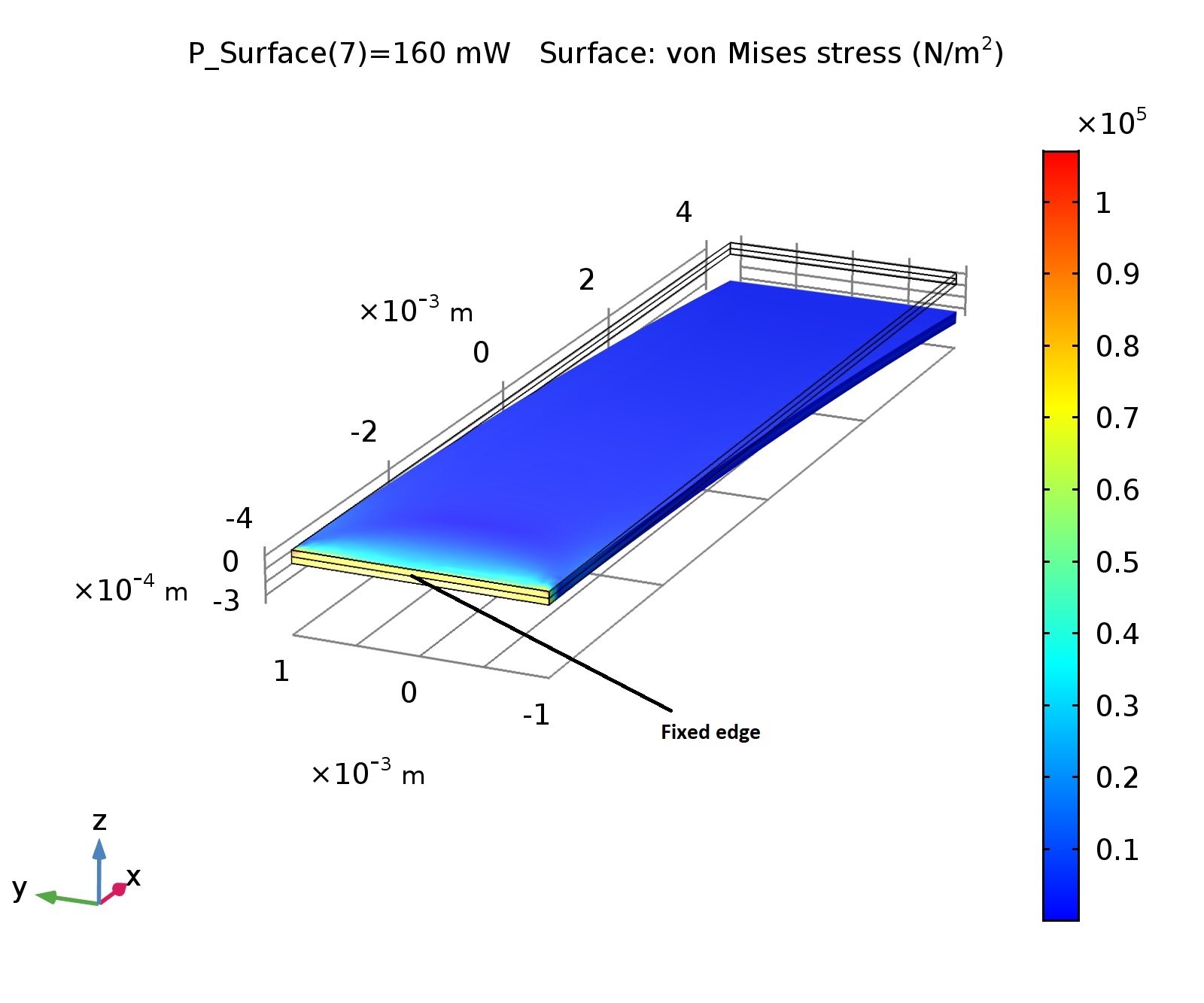}
      \includegraphics[scale = 0.57]{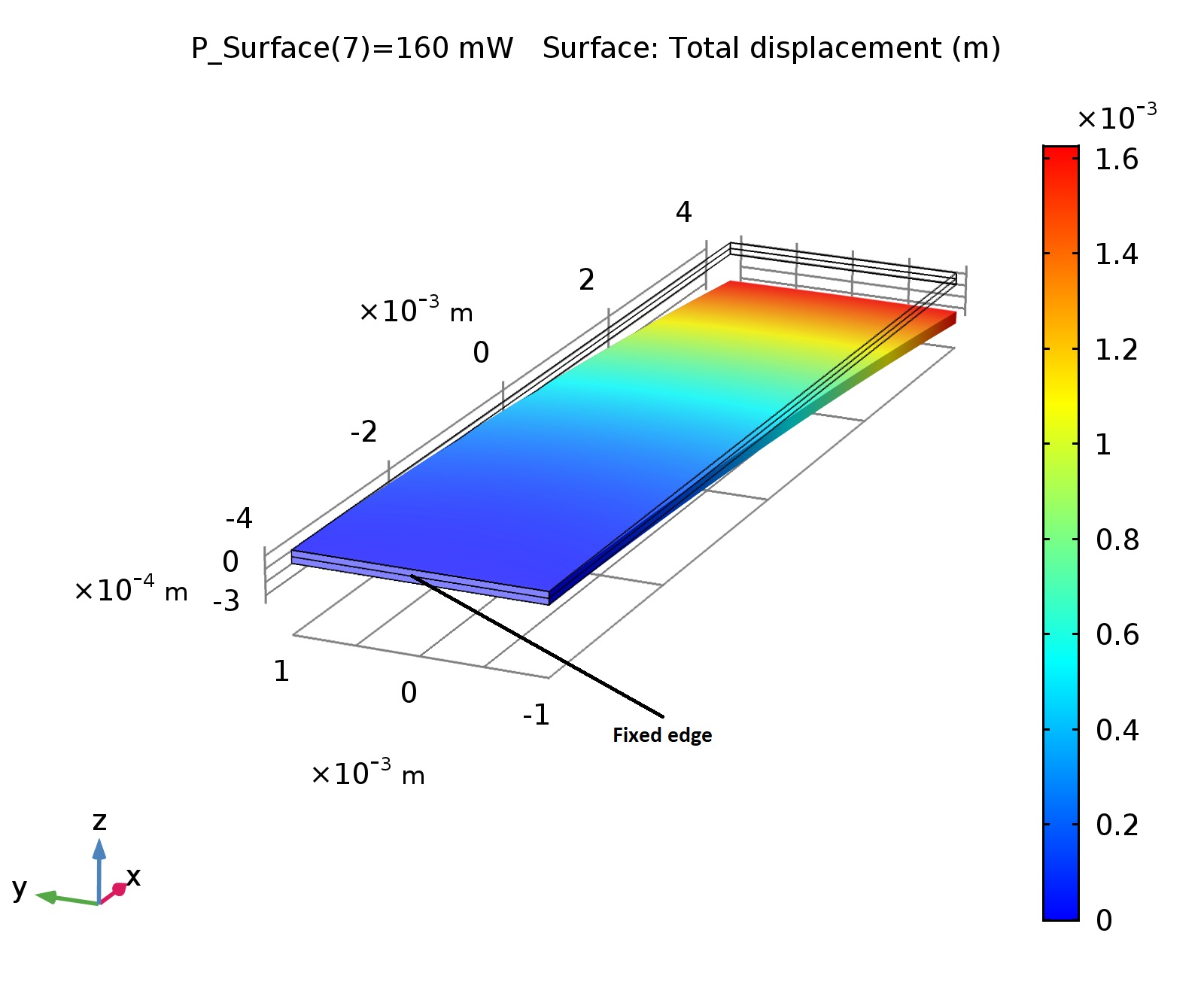}
   \end{tabular}
   \end{center}
   \caption[4] 
   { \label{fig:4} 
Thermal profile at the C-Side when $P_{surface} = 160$ \si{mW} (top left). The temperature at the tip is 389.483 \si{K}. Isothermal contours showing different temperature regions (top right). von Mises stress profile (bottom left), and total deflection (bottom right) when $P_{surface} = 160$ \si{mW}. The deflection is towards the C-Side and the laser source and away from the face of the magnet.}
   \end{figure}
      \begin{figure} [ht]
   \begin{center}
   \begin{tabular}{c} 
   \includegraphics[scale = 0.57]{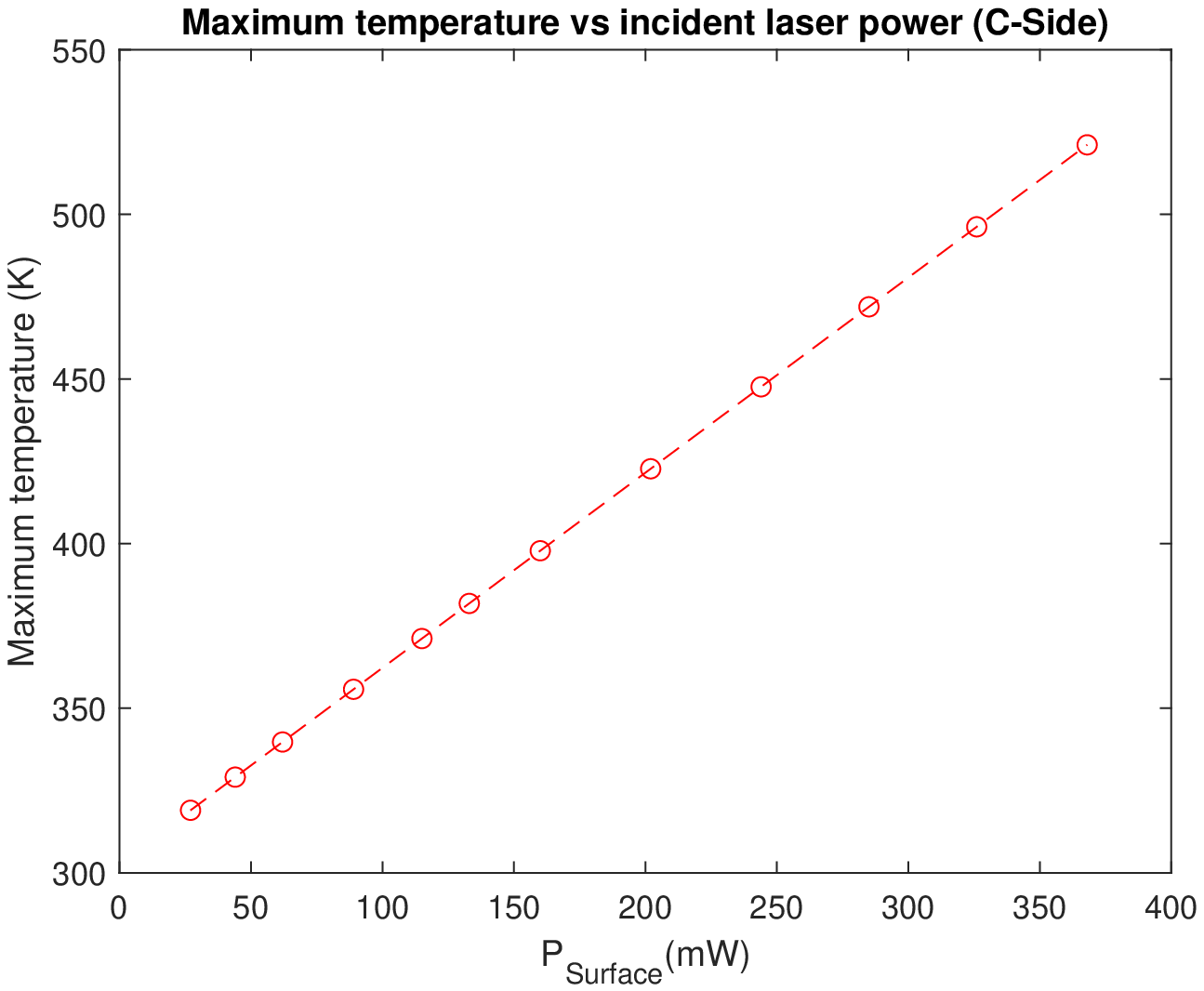}
     \includegraphics[scale = 0.57]{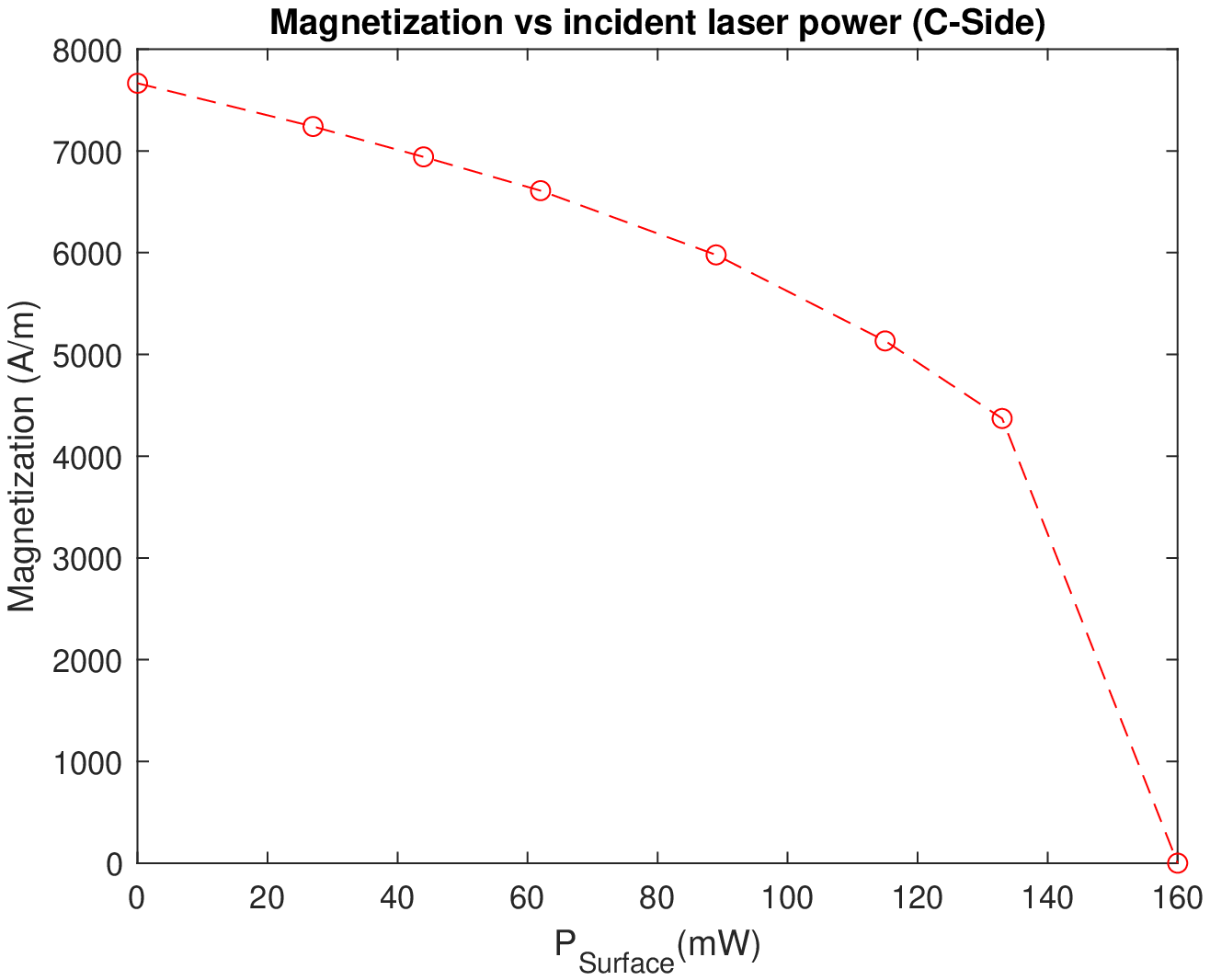}
     \\
   \end{tabular}
   \end{center}
   \caption[5] 
   { \label{fig:5} 
Maximum temperature at the C-Side (left) and magnetization value at the C-Side (right) of the composite structure at different incident laser power ($P_{surface}$). }
   \end{figure}
      \begin{figure} [ht]
   \begin{center}
   \begin{tabular}{c} 
   \\
   \includegraphics[scale = 0.57]{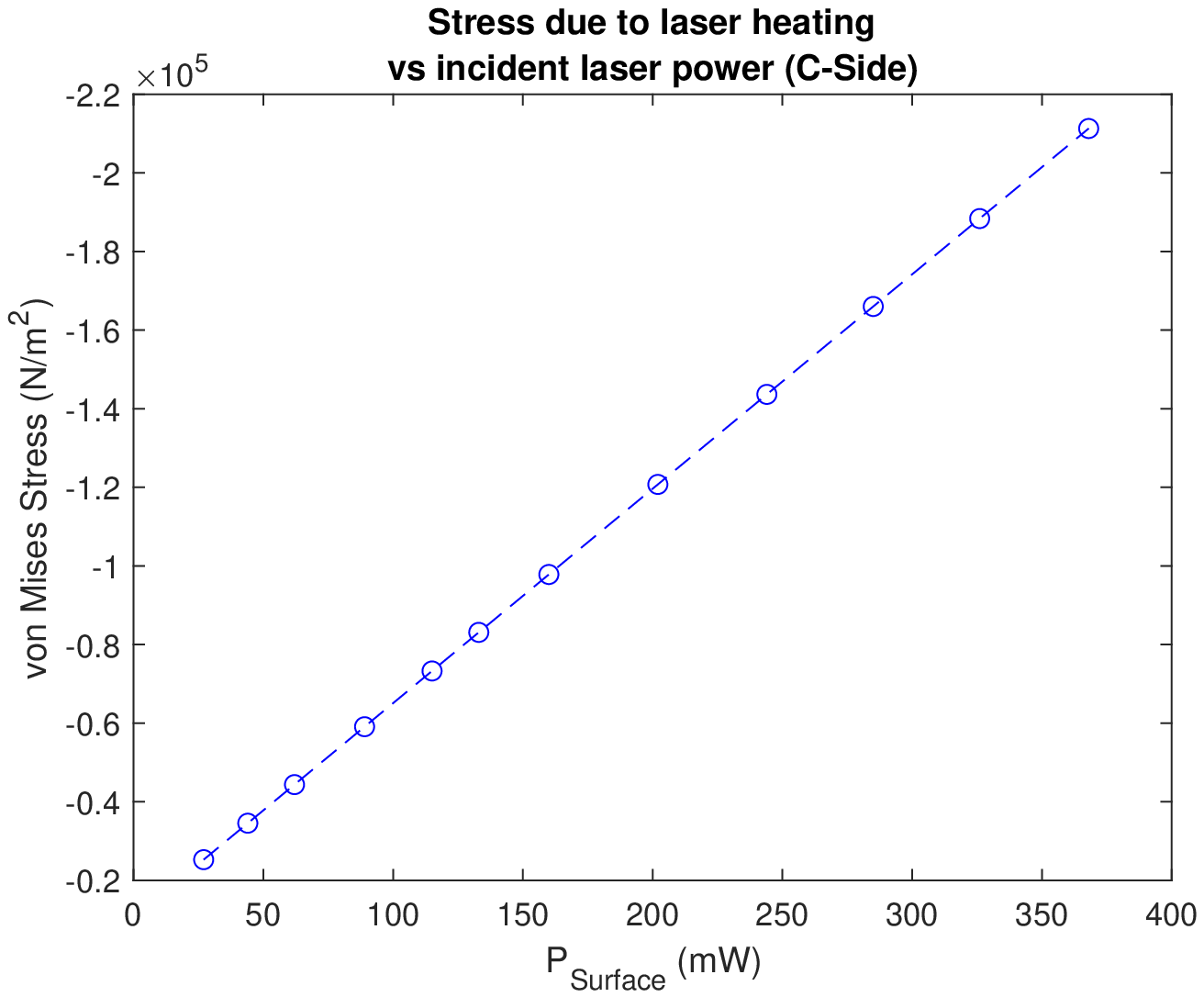}
     \includegraphics[scale = 0.57]{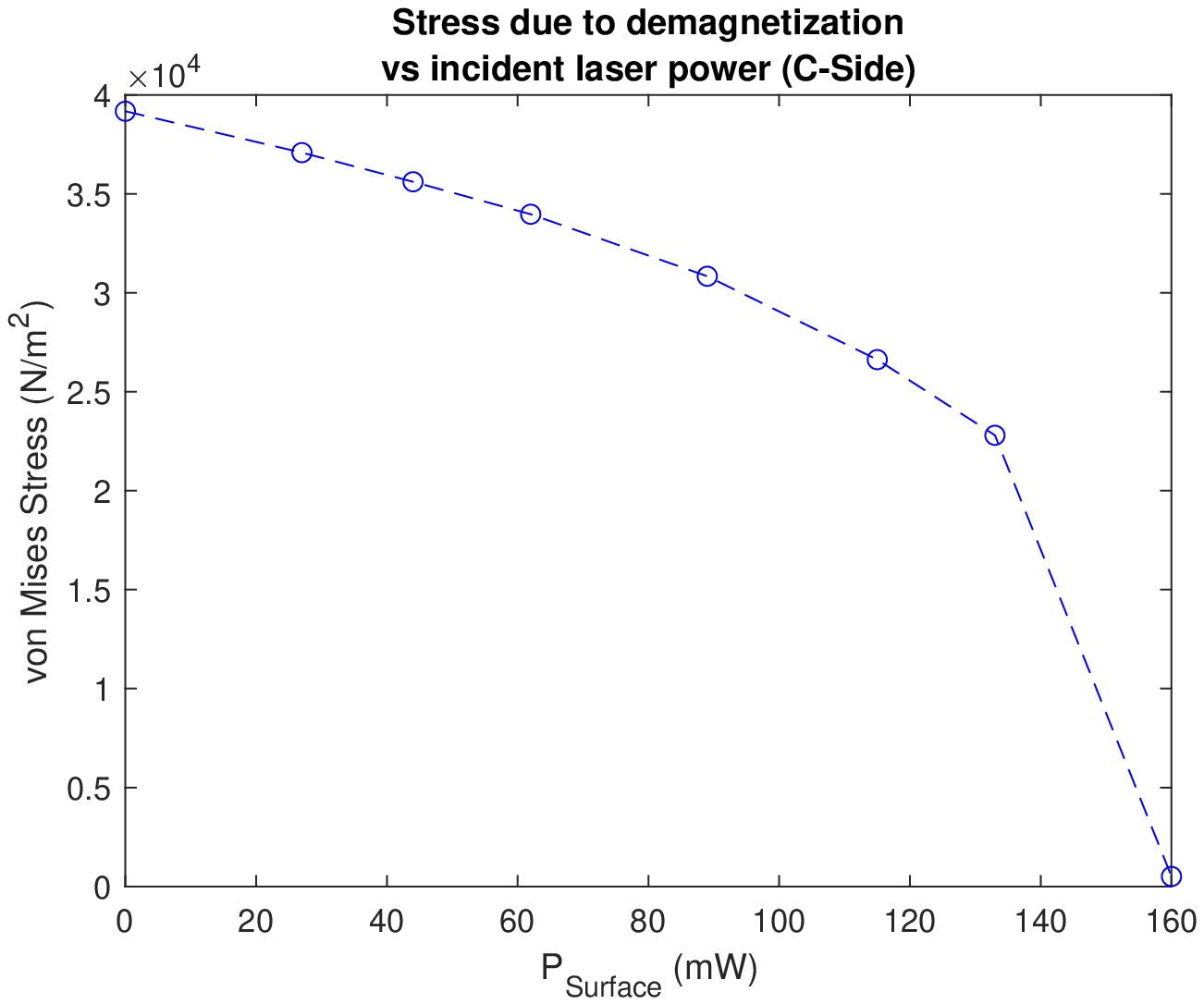}
     \\
   \end{tabular}
   \end{center}
   \caption[6] 
   { \label{fig:6} 
Maximum von Mises stress due to laser heating (left) and due to demagnetization effect (right) of the composite structure at different incident laser power ($P_{surface}$). }
   \end{figure}
      \begin{figure} [ht]
   \begin{center}
   \begin{tabular}{c} 
   \\
   \includegraphics[scale = 0.57]{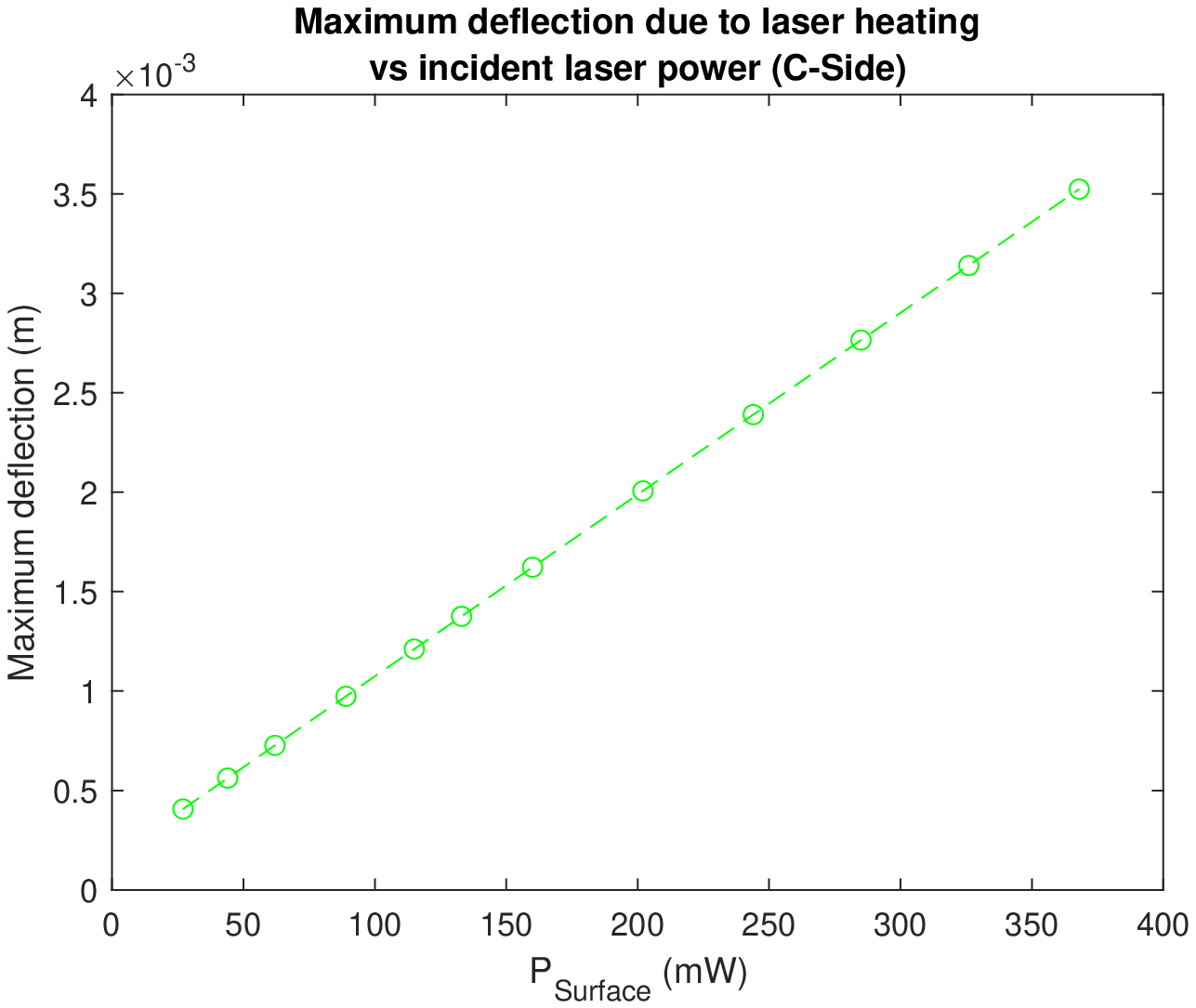}
     \includegraphics[scale = 0.57]{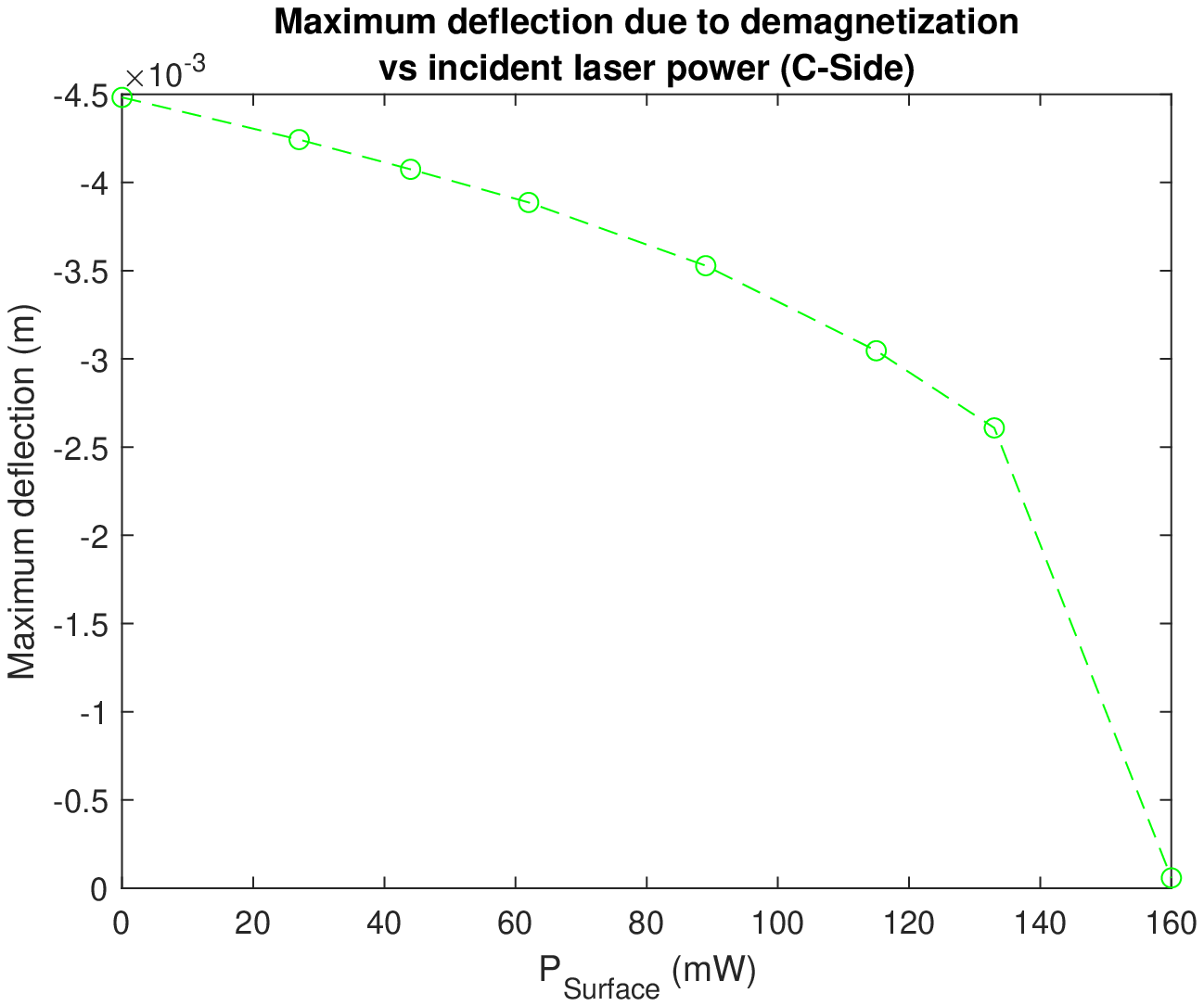}
     \\
   \end{tabular}
   \end{center}
   \caption[7] 
   { \label{fig:7} 
Maximum deflection due to laser heating (left) and due to demagnetization effect (right) of the composite structure at different incident laser power ($P_{surface}$). The deflection due to laser heating is towards C-Side while the magnetic load is towards the P-Side.}
   \end{figure}
   \begin{figure}[ht]
   \begin{center}
   \begin{tabular}{c}
   \\
   \includegraphics[scale = 0.57]{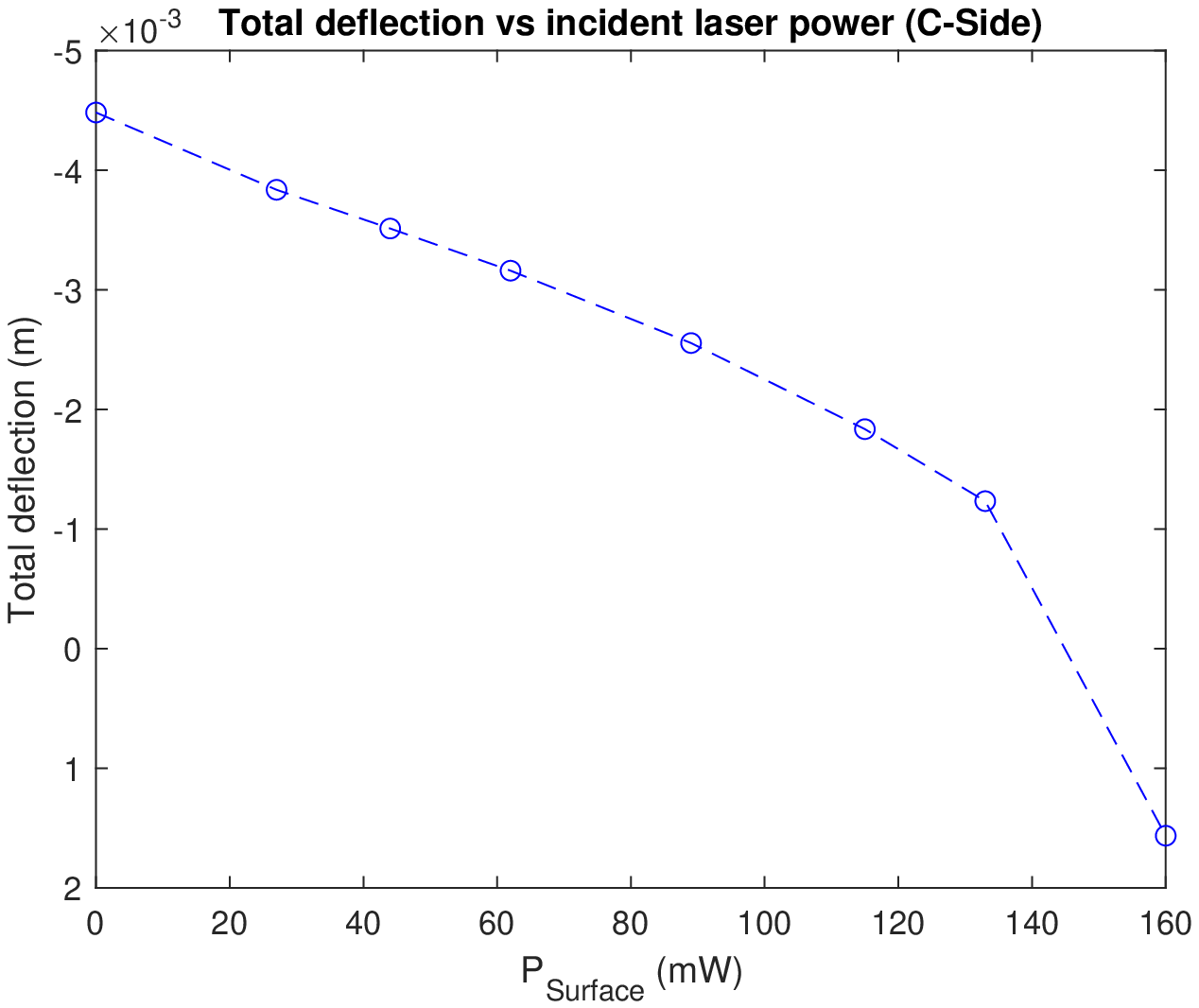}
   \includegraphics[scale = 0.57]{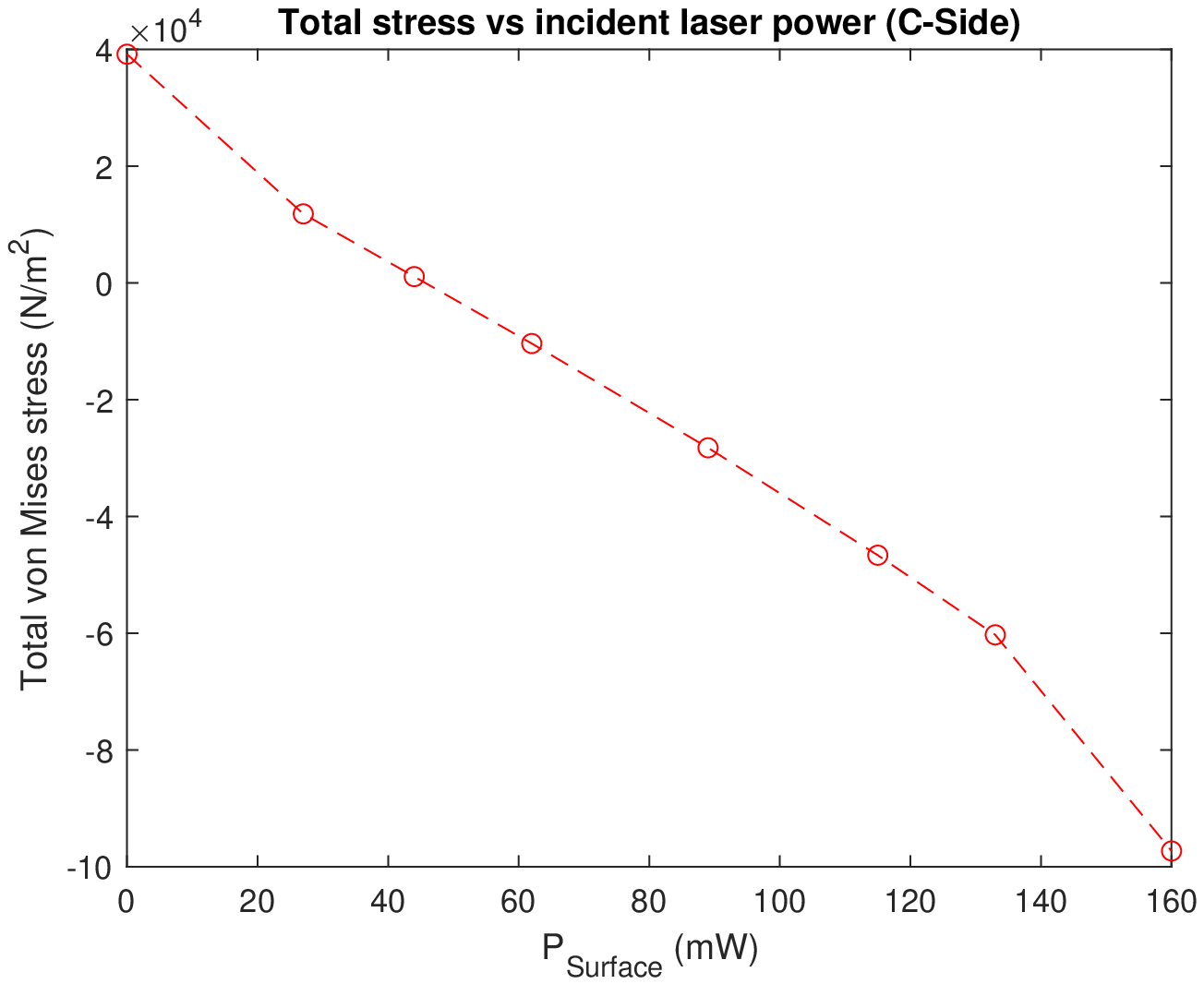}
   \end{tabular}
   \end{center}
   \caption[8]
   {\label{fig:8}
   Total deflection due to both laser heating and magnetic loading at different $P_{surface}$ (left). Total stress due to both laser heating and magnetic loading at different $P_{surface}$ (right). The direction of deflection is towards the P-Side at first and gradually changes towards the C-Side due to laser heating and demagnetization effects. The stress also shows the same behavior but in opposite direction as it counters the deflection.  }
   \end{figure}
   \begin{figure}[ht]
   \begin{center}
   \begin{tabular}{c}
   \\
   \includegraphics[scale = 0.55]{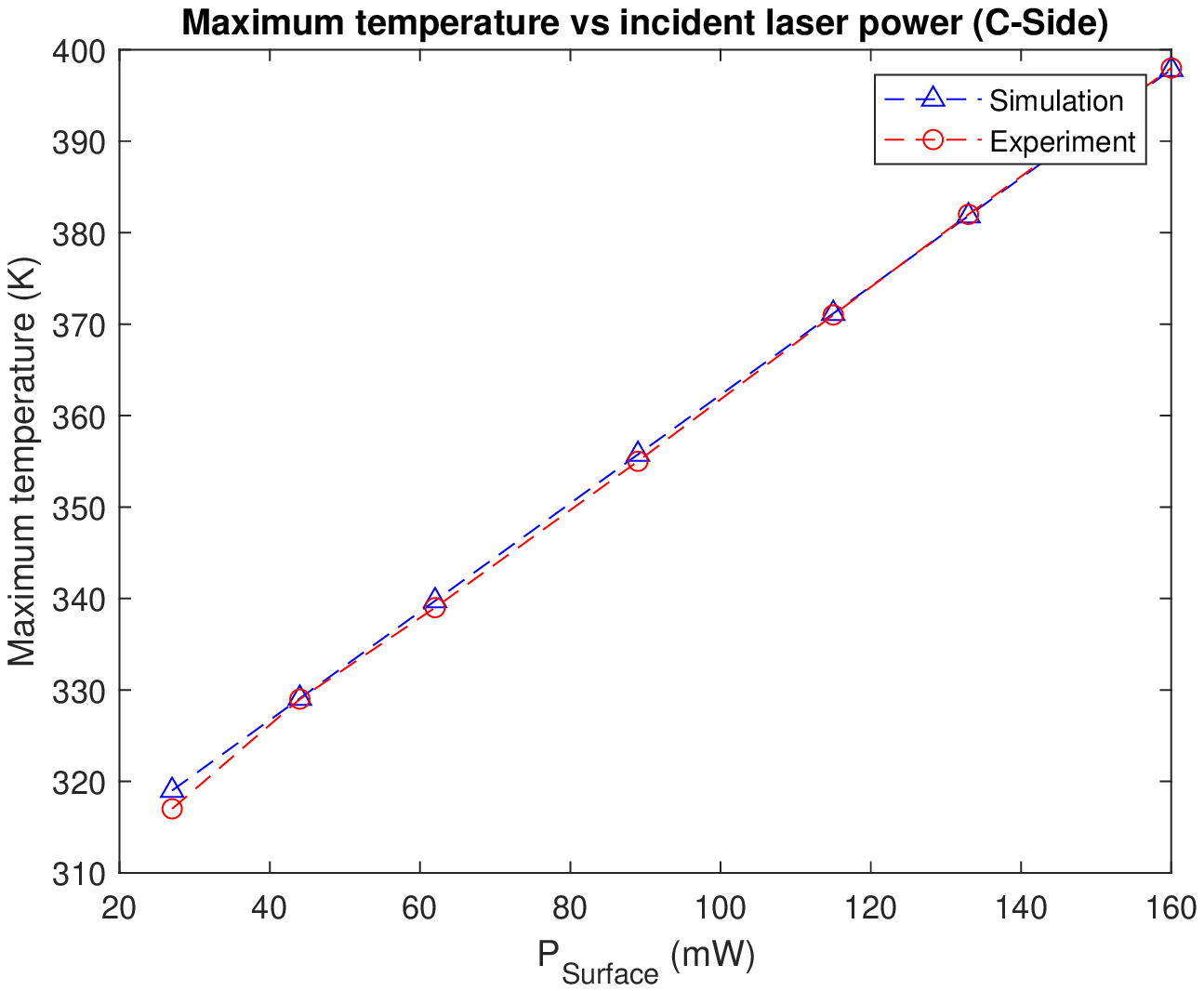}
   \includegraphics[scale = 0.55]{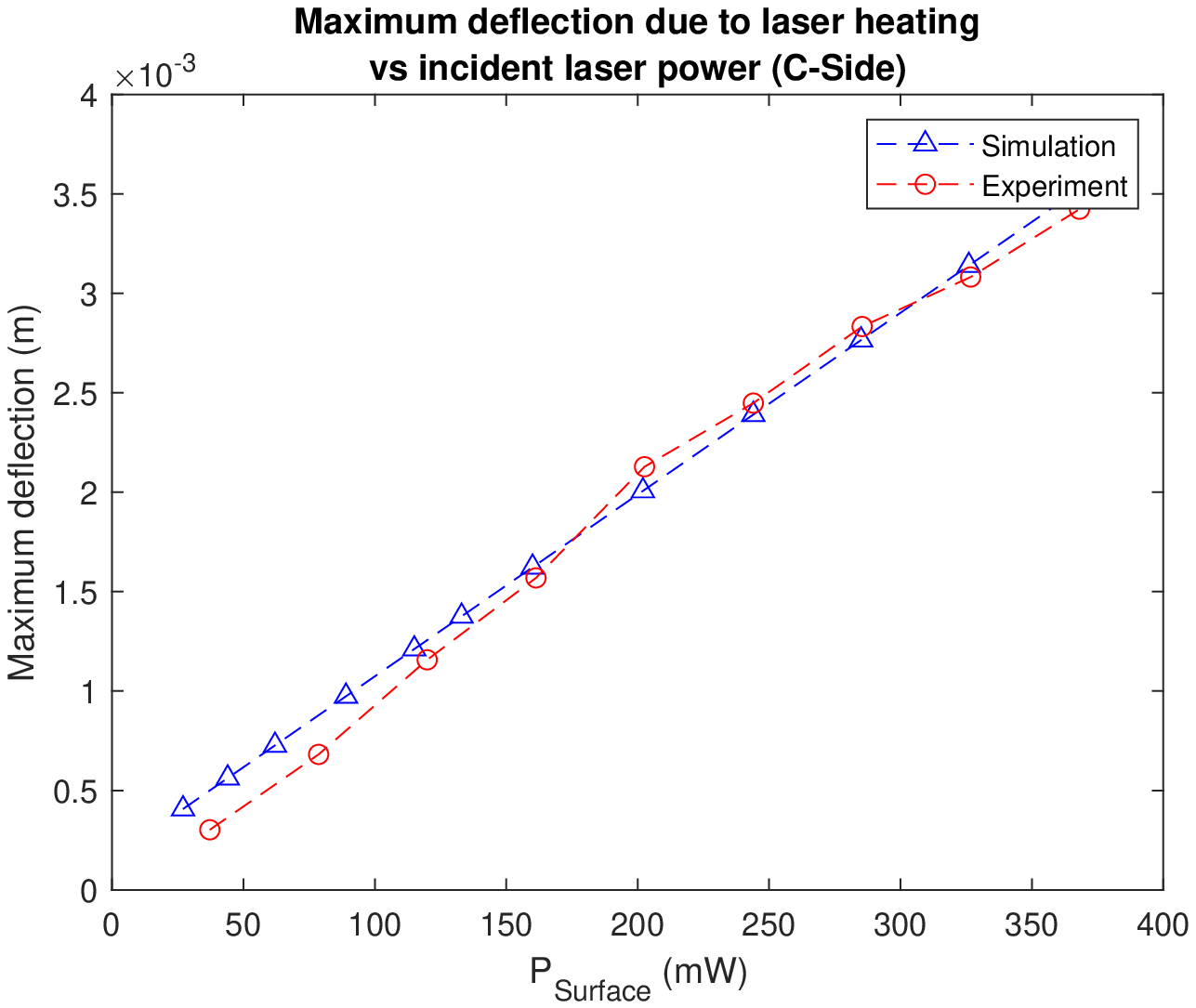}
   \\
   \includegraphics[scale = 0.55]{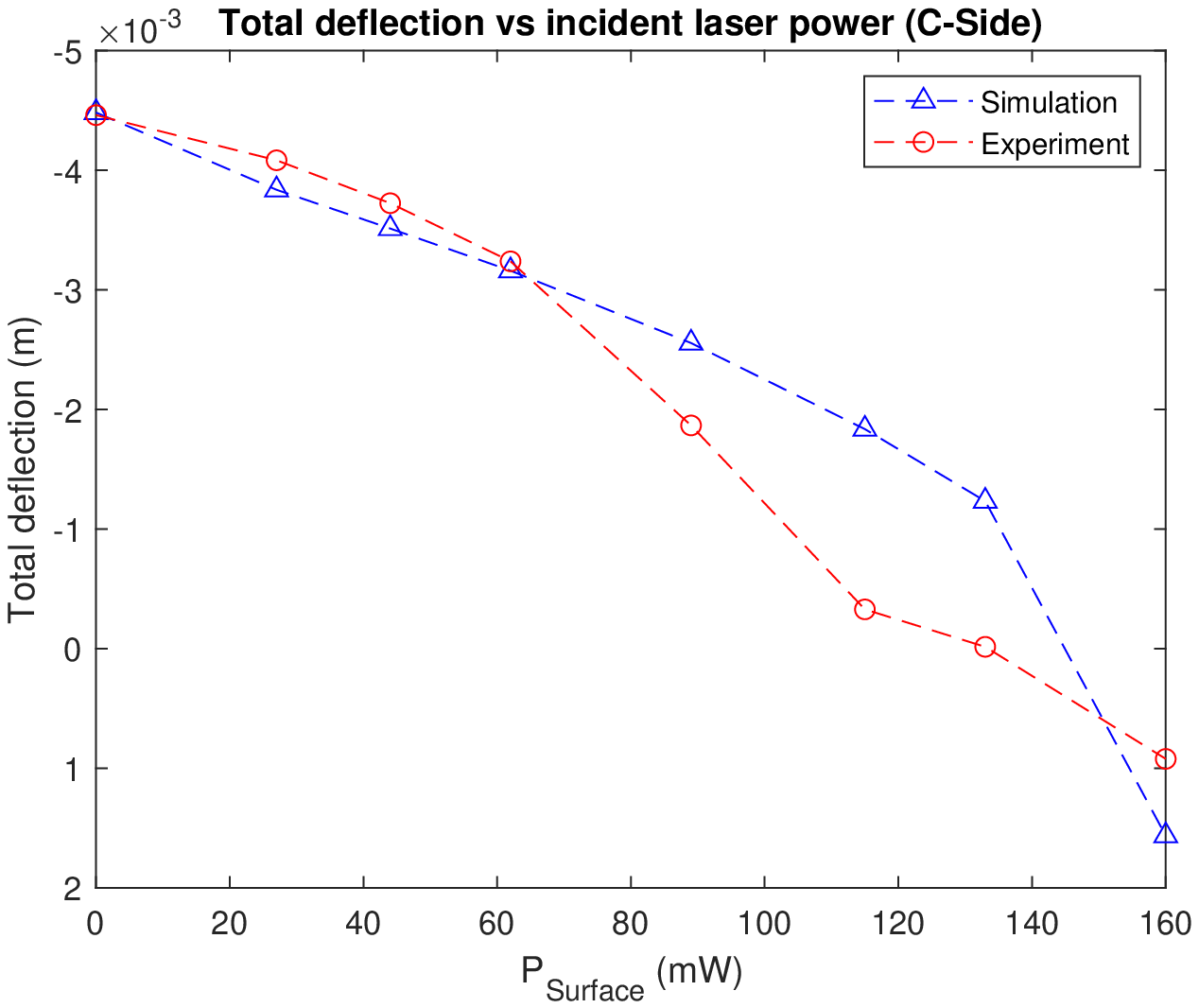}
   \end{tabular}
   \end{center}
   \caption[9]
   {\label{fig:9}
   Comparison between simulation results obtained from COMSOL and experimentally obtained data from Meng Li et al.~\cite{li2018flexible} Maximum surface temperature due to laser heating at the C-Side (top left). Maximum deflection due to laser heating at the C-Side (top right). Total deflection due to demagnetization and laser heating at the C-Side (bottom center).}
   \end{figure}
  Figure~\ref{fig:d1} is an illustration describing the direction of deflection of the composite structure under different stimuli. The composite structure deflects towards the P-Side when only magnetic load is applied from the P-Side. However, the direction of deflection is towards the C-Side when a laser heating source is used. Finally, when both the magnet and the laser source is used, the deflection happens from P-Side to C-Side. We are following a convention where the deflection towards the P-Side is considered as -ve and towards the C-Side as +ve. The same convention is also followed for the stress generated in the composite structure as well. 
  \\
  Figure~\ref{fig:3} shows the magnetic field lines (in grey) and magnetic flux density generated at the P-Side of the composite geometry. We can see that the model can simulate a 30 \si{mT} magnetic flux density at the P-Side of our composite structural model. It can also be seen that the magnetic force acting on the structure is attractive and the deflection direction is towards the face of the magnet. The stress profile has also been shown in fig~\ref{fig:3}. A maximum deflection of 4.48 \si{mm} is observed when there is no laser power incident on the C-Side.
  \\
  Figure~\ref{fig:4} shows the simulation results for laser heating at the C-Side in the absence of any magnetic field source. It can be seen that the surface temperature reaches to nearly 400 \si{K} when the incident laser power at the C-Side is about 160 \si{mW}. The temperature distribution and isothermal contours validate the gaussian nature of the heat source. Due to laser heating, thermal stress and deflection are generated which can be seen in the 3D plots included in fig~\ref{fig:4}. A maximum deflection of 1.6 \si{mm} is observed towards the direction of the laser heating source (C-Side) when $P_{Surface}=160$ \si{mW}. The direction of deflection is opposite to that of the deflection occurring due to the magnetic load.
  \\
  Figure~\ref{fig:5} shows the maximum temperature and change in the magnetization value of the composite structural model. With increasing incident laser power, the temperature of the model increases linearly causing demagnetization which is validated by the plot as shown in fig~\ref{fig:5}.
 \\
  Figure~\ref{fig:6} and fig~\ref{fig:7} show the maximum stress and deflection generated in our composite structural model due to laser heating and magnetic forcing respectively. The direction of deflection due to magnetic forcing is towards the P-Side whereas the direction of deflection due to laser heating is towards the C-Side. With increasing incident laser power, thermal stress and thermal deflection in the system increases while the stress and deflection due to magnetic forcing decreases due to demagnetization.
 \\
 In fig~\ref{fig:8}, we have plotted the total stress and total deflection occurring towards the P-Side taking into account both laser heating and demagnetization effects. As the incident laser power increases, demagnetization effects dominate in the system causing generation of overall lesser stress and deflection.
 \\
 Figure~\ref{fig:9} shows the comparison between the simulation results obtained from COMSOL model and the experimental data obtained from Meng Li et al.~\cite{li2018flexible}. We have shown the maximum surface temperature and maximum deflection obtained due to the laser heating at the C-Side of the composite structure. Figure~\ref{fig:9} also includes comparison results for the total deflection due to demagnetization and laser heating of the composite structure.
\section{Conclusion}
\label{sec:4}
From the simulation results, we can conclude that the model successfully simulates the laser heating and demagnetization effects of our composite layered structure. A deflection of 4.48 \si{mm} is observed when the laser is off at 303 \si{K} (room temperature) which tends to gradually decrease with increasing incident laser power at the C-Side of our composite structure. Due to demagnetization and laser heating, the structure further deflects to 1.6 \si{mm} towards the C-Side, thus, accounting for a total displacement of 6.08 \si{mm}. The nature of deflection is similar to what has been reported in the works of Meng Li et al.~\cite{li2018flexible} and the thermal and demagnetization effects also validate with their work (fig~\ref{fig:9}). We also observed some offset between the simulation and the experimental data in fig~\ref{fig:9} which can be due to the sensitive dependence of the  laser heating and demagnetization model on the model parameters which can be further explored in future work.
\\
Hence, our FEM simulation model successfully reproduces the previously observed motion of the ($\text{CrO}_\text{2}$+PDMS) structure and one can use our modeling approach in a light controlled actuated structures with varying material properties and dimensions. The advantage of our simulation model is that it also offers information about the von Mises stress generated in the composite structure, thus, helping in performing fracture and reliability analysis. 
 \section*{Acknowledgements}       
The authors would like to thank Masud Mansuripur, Professor at the Wyant College of Optical Sciences, University of Arizona and all the members of the University of Arizona Space Astrophysics (UASAL) Lab for their constructive suggestions. This work was supported by the University of Arizona Technology and Research Initiative Fund (TRIF).

\end{document}